\pdfoutput=1	

\documentclass[reprint, superscriptaddress, amsmath,amssymb, aps]{revtex4-1}

\usepackage[utf8]{inputenc}
\usepackage[english]{babel}

\usepackage{color} 
\usepackage{graphicx}

\usepackage{etoolbox}	

\usepackage[linesnumbered,ruled]{algorithm2e}
\SetAlgoCaptionSeparator{}	
\makeatletter
\newenvironment{namedalgorithm}[1]{%
	\medskip
	\renewcommand{\algocf@algocfref}{#1}
	\SetNlSty{textbf}{}{\enspace} 
	\DontPrintSemicolon 
	\begin{algorithm}[H] %
	}{\end{algorithm}%
	\medskip}
\makeatother

\usepackage{dcolumn} 
\usepackage[inline]{enumitem} 

\usepackage{amsmath}
\usepackage{bm}      	
\usepackage{braket}

\usepackage{hyperref}	

\newcommand{\RSVDPREFIX}{R}
\newcommand{\DSVDPREFIX}{T}

\newcommand{\RSVD}{\RSVDPREFIX{}SVD}
\newcommand{\DSVD}{\DSVDPREFIX{}SVD}

\newcommand{\bTTN}{TTN}

\newcommand{\precision}{\delta E}
\newcommand{\bonddim}{\chi}
\newcommand{\localdim}{d}
\newcommand{\field}{h}
\newcommand{\fieldcrit}{h_c}
\newcommand{\timestep}{dt}   
\newcommand{\speedup}{\tau}

\newcommand*{\complexity}{\mathcal{O}}

\newcommand*{\globaloptr}{}

\newcommand*{\dtrunc}{\delta_\text{trunc}}

\newcommand*{\nsectors}{N}

\begin{document}


\title{Probabilistic low-rank factorization accelerates tensor network simulations of critical quantum many-body ground states}

\author{Lucas Kohn}
\affiliation{Institute for Complex Quantum Systems and Center for Integrated Quantum Science and Technologies, Universit\"at Ulm, 89069 Ulm, Germany}

\author{Ferdinand Tschirsich}
\email{ferdinand.tschirsich@uni-ulm.de}
\affiliation{Institute for Complex Quantum Systems and Center for Integrated Quantum Science and Technologies, Universit\"at Ulm, 89069 Ulm, Germany}

\author{Maximilian Keck}
\affiliation{NEST, Scuola Normale Superiore and Istituto Nanoscienze-CNR, 56126 Pisa, Italy}

\author{Martin B. Plenio}
\affiliation{Institute for Theoretical Physics, Universit\"at Ulm, 89069 Ulm, Germany}

\author{Dario Tamascelli}
\affiliation{Institute for Theoretical Physics, Universit\"at Ulm, 89069 Ulm, Germany}
\affiliation{Dipartimento di Fisica, Università degli Studi di Milano, 20133 Milano, Italy}

\author{Simone Montangero}
\affiliation{Institute for Complex Quantum Systems and Center for Integrated Quantum Science and Technologies, Universit\"at Ulm, 89069 Ulm, Germany}
\affiliation{Theoretische Physik, Universit\"at des Saarlandes, 66123 Saarbr\"ucken, Germany}
\affiliation{Dipartimento di Fisica e Astronomia, Università degli Studi di Padova, 35131 Padova, Italy}

\date{\today}


\begin{abstract}
We provide evidence that randomized low-rank factorization is a powerful tool for the determination of the ground state properties of low-dimensional lattice Hamiltonians through tensor network techniques. 
In particular, we show that randomized matrix factorization outperforms truncated singular value decomposition based on state-of-the-art deterministic routines in TEBD and DMRG-style simulations, even when the system under study gets close to a phase transition:
We report linear speedups in the bond- or local dimension, of up to 24 times in quasi-2D cylindrical systems.
\end{abstract}



\maketitle




\section{Introduction}

Tensor network (TN) methods have long proven their power as an indispensable tool in simulating quantum- and classical many-body systems \cite{Schollwoeck2011DMRG_MPS_review,Orus2014TN_review}.
As first realized by S.\ White with the density matrix renormalization group (DMRG) algorithm \cite{White1992DMRG,White1993DMRG}, a variational ansatz on the manifold of matrix product states (MPS) \cite{Rommer1997MPS}, TNs provide an efficient parameterization of low-entangled wave functions in quantum many body state-space \cite{Eisert2013TNReview_Entanglement}.
While the MPS naturally captures the relevant low-energy spectrum in particular of one-dimensional (1D) gapped Hamiltonians obeying area laws of entanglement \cite{Audenaert2002EntanglementChain,Plenio2005AreaLaw, Eisert2010AreaLaw,Wolf2006AreaLawFermions}, TNs have been generalized to more complex scenarios:
In over two decades of evolution, they have been successfully applied to higher dimensions \cite{Verstraete2006PEPS,Tagliacozzo2009TTN2D,Gerster2017bTTNHofstadter}, critical phenomena \cite{Vidal2007MERA,Vidal2008MERA,Silvi2010bTTN,Orus2014MultipartiteTN}, as well as finite temperature and the study of closed and open system dynamics \cite{Verstraete2004MPDO,Werner2016LPTN}, and lattice gauge theories \cite{Silvi2014LatticeGaugeTN,Tagliacozzo2014LatticeGaugeTN,Pichler2016U1LatticeGauge}, just to name a few examples.
TNs have also been equipped with structure to encode and exploit symmetries in the model under investigation \cite{Singh2010SymTN,Singh2011SymU1,Singh2012SymSU2,Weichselbaum2012SymNonabelian}.

Truncated singular value decompositions (SVDs) are widely used in TN algorithms to compress states into their respective TN state manifold.
Examples include the time-evolving block decimation (TEBD) \cite{Vidal2003TEBD,Vidal2004TEBD}, the tensor renormalization group \cite{Levin2007TRG}, the corner transfer matrix renormalization group \cite{Nishino1996CTMRG} and the projected entangled pair states \cite{Verstraete2004PEPS, Murg2007PEPS2}, but also traditional DMRG which is often formulated in terms of truncated eigenvalue decomposition.
In TN numerical practice, SVDs have the additional advantage to provide relevant isometries by orthonormality of the singular vectors, and reveal valuable information of the encoded network state, e.g.\ in the form of entanglement measures based on singular values. 

The traditional way to compute a truncated SVD is to first perform the full SVD of a matrix, and then discard the smallest singular values.
This is reliable and accurate, but also a very costly operation that often dominates computational complexity of TN algorithms.
Intuitively, it is also not the most economic protocol: 
A lot of effort is spent in computing all singular values and -vectors, many of which are then discarded.
By avoiding the full SVD, a truncated SVD can be obtained more efficiently, especially when the number of retained singular values is small.
Well known methods of this class are simultaneous subspace iteration or Krylov subspace methods like Lanczos- or implicitly restarted Arnoldi algorithms \cite{Golub2012MatrixComputations,Demmel1997NumericalLA}. 
Their relevance in large-scale data classification and compression in `big data' applications \cite{Erichson2017RandomCPTensorDecomp,Hastie2009StatisticalLearning}, signal processing \cite{Cichocki2015SignalProcessingTensors}, face recognition \cite{Vijayakumari2013FaceRecognitionSurvey,Pai2015FaceRecognitionIllumination}, DNA analysis \cite{Wall2003MicroarrayAnalysisSVD} and other fields is a driving force behind the ongoing development of faster algorithms.
A use case in the approximative contraction of unstructured TNs has also been reported \cite{Jermyn2017automatic}.

Randomized algorithms outperform prior approaches in both speed and reliability \cite{Halko2011LowRankProbabilistic}. 
Specifically, the randomized SVD (\RSVD{}) based on a probabilistic low-rank matrix-factorization algorithm \cite{Halko2011LowRankProbabilistic} is capable of delivering accurate results with failure probabilities that can be made arbitrarily small, independent of peculiar choices like starting vectors that are common in deterministic methods. 
\RSVD{} thus promises to significantly accelerate TN methods that spend a considerable amount of resources in truncated SVDs.

Recently, significant speedup due to \RSVD{} has been reported in the TEBD simulation of open system dynamics \cite{Tamascelli2015TEBDRSVD}. 
In particular, the authors of \cite{Tamascelli2015TEBDRSVD} showed that the robust \RSVD{} outperforms deterministic SVD algorithms in delivering a limited number of largest singular values (and corresponding vectors) while maintaining high accuracy in the simulated dynamics. 
It is however an open question whether \RSVD{} can be applied  with similar success in scenarios beyond the open system dynamics, since \RSVD{} performance and accuracy are closely tied to the encountered spectra of singular values.
This question applies especially to critical systems where the singular values are expected to decay slowly.

In this paper we demonstrate superior performance of \RSVD{} in the very original application field of TN methods, namely in identifying ground state properties of low-dimensional quantum lattice Hamiltonians. 
We confirm significant speedup in different physical scenarios, including situations when the system is critical.
Embedded in full-fledged TN simulations, we compare the \RSVD{} against the truncated full SVD from state-of-the-art LAPACK implementations \textit{D/ZGESDD} \cite{Anderson1999LAPACK_userguide}, referred to as \DSVD{} in the following. 
As benchmarks, we use variants of the quantum Ising model in imaginary TEBD time evolution and a DMRG-style ground state search with the hierarchical binary tree TN (\bTTN{}) \cite{Shi2006TTN,Hackbusch2009TTNScheme,Silvi2010bTTN,Murg2010TTN,Nakatani2013TTN,Gerster2014TTN,Gerster2016BoseHubbard}.
It will become apparent that a simple replacement of \DSVD{} with \RSVD{}-code can lead to speedups between one and two orders of magnitude, while preserving the same precision, even when state-of-the-art TN techniques are employed \cite{Singh2010SymTN,Singh2011SymU1}.

The paper is organized as follows: 
First, we motivate the use of truncated SVD as a tool of information compression in typical TN scenarios in Sec.~\ref{sec:Compression}. 
We continue with a short review of the \RSVD{} method and how it can help achieving faster compression in
Sec.~\ref{sub:RSVDAlgorithm}. 
We then introduce our benchmark models in Sec.~\ref{sec:BenchmarkSetup} and present a detailed performance analysis by switching from \DSVD{} to \RSVD{} in Sec.~\ref{sec:Results}.
Sec.~\ref{sec:Discussion} concludes the paper with a discussion of the results and with practical tips for the implementation and identification of situations that may benefit from \RSVD{}.

\section{Low-rank factorization}\label{sec:Compression}

The maximal bond dimension $\bonddim$ of a TN is a fundamental parameter: 
It can be linked to the amount of quantum entanglement that can be hosted in the network state \cite{Eisert2013TNReview_Entanglement}. 
At the same time, $\bonddim$ determines the computational complexity of algorithms performed on the network. 
Typical operations include the computation of expectation values, propagation in real or imaginary time and renormalization-steps updating the network description in iterative algorithms.
All these operations can result in the growth of index dimensions beyond the maximally allowed bond dimension. 
A compression step is then achieved by means of a truncated SVD.

\subsection{Truncated SVD}
Let $A$ be a real- or complex-valued $m$-by-$n$ matrix with $m\ge n$.
In our case, $A$ usually represents the contraction of two tensors, and it can also be given in the form of a matrix product $X'Y'$. 
The compression step then provides a rank-$\bonddim$ factorization $XY$ which is a good approximation $A\approx XY$, but also limits $X$ to an $m$-by-$\bonddim$ matrix and $Y$ to a $\bonddim$-by-$n$ matrix.
A standard solution is to compute the rank-$\bonddim$ truncated SVD as follows:

\begin{namedalgorithm}{\DSVD}
\caption{} 
\label{alg:truncatedSVD}
	\KwIn{$m$-by-$n$ matrix $A$, integer $\bonddim$}
	\KwOut{rank-$\bonddim$ truncated SVD of $A$}
	\BlankLine
	Compute the full SVD of $A=U \Sigma V^\dagger$\;
	Extract the $\bonddim$ largest singular values from $\Sigma$ and corresponding columns of $U$ and $V$\;
\end{namedalgorithm} 

In particular, $U$ is an $m$-by-$n$ matrix, $\Sigma$ and $V$ are $n$-by-$n$ matrices, and we assume $\Sigma$ is the diagonal matrix containing the singular values $\sigma_j=\Sigma_{jj}$ in descending order  $\sigma_1\ge\sigma_2\ge\dots\ge\sigma_n\ge0$. 
We then discard the $n-\chi$ smallest singular values (assuming $\bonddim \le n$) and obtain for instance $X_{ij}=U_{ij}$ and $Y_{jk}=\sigma_j V^\dagger_{jk}$ for $j=1,\dots,\bonddim$.

The truncation error $\dtrunc:=\|A-XY\|$ is then known to be minimal \cite{Eckart1936LowRank,Mirsky1960Symmetric} when measured in spectral norm ($\dtrunc=\sigma_{\bonddim+1}$) or Frobenius norm ($\dtrunc^2=\sum_{k=\bonddim+1}^{n} \sigma_k^2$).

The availability of highly optimized SVD routines makes the implementation of \DSVD{} straightforward. 
However, while it provides high accuracy, actually computing all $n$ singular values and -vectors in the full SVD of $A$ still requires $\complexity(mn^2)$ floating-point operations.

When the compression ratio $\bonddim/n$ becomes small, a more efficient protocol for computing the truncated SVD of $A$ is the \RSVD{} algorithm. 

\subsection{Randomized algorithm}\label{sub:RSVDAlgorithm}

The basic idea of \RSVD{} is simple: 
First, the input matrix $A$ is approximated with a rank-$\ell$ matrix $A_\ell \approx A$, which is obtained with randomness. From there, a rank-$\bonddim$ truncated SVD of $A_\ell$ is obtained at significantly lowered computational cost compared to a full SVD of $A$.

Two characteristic choices lead to an accurate $A_\ell$: 
\begin{enumerate*}
	\item Oversampling the approximation with $\ell>\bonddim$  \cite{Martinsson2011RandLowRankNoPI}, and
	\item employing a randomized power-iteration  of length $q$ \cite{Rokhlin2010LowRankPIScheme}.
\end{enumerate*}
We state the complete algorithm first, as put forward in \cite{Halko2011LowRankProbabilistic}, and then discuss the impact of both parameters $\ell$ and $q$ on computational cost and quality of the outcome.

\begin{namedalgorithm}{\RSVD{}}
	\caption{} 
	\label{alg:randomizedSVD}
	\KwIn{$m$-by-$n$ matrix $A$, integers $\bonddim$, $\ell$, $q$}
	\KwOut{approximate rank-$\bonddim$ truncated SVD of $A$}
	\BlankLine
	Generate an $n$-by-$\ell$ Gaussian matrix $\Omega$\;\label{alg:RSVDgaussian}
	Compute $Y:=\left(AA^\dagger\right)^q A\Omega$\;\label{alg:RSVDsample}
	Store in $Q$ the orthonormalized columns of $Y$\;\label{alg:RSVDortho}
	Compute the rank-$\bonddim$ truncated SVD of $B:=Q^\dagger A$\;\label{alg:RSVDtruncate}
\end{namedalgorithm}

In detail, the algorithm begins by drawing a random test matrix $\Omega$ from a standard Gaussian distribution in step~\ref{alg:RSVDgaussian}. 
Note that other choices may work as well, and that the quality of random numbers is not of crucial importance.
Step~\ref{alg:RSVDsample} then produces an $m$-by-$\ell$ sample $Y$ of the range of $A$, by multiplying the columns of the test-matrix with $\left(AA^\dagger\right)^q A$. 
This process emphasizes the most relevant singular vectors, associated with large singular values $\sigma$, by a factor of $\sigma^{2q+1}$. 
In order to maintain numerical stability when these factors range over several orders of magnitude, step~\ref{alg:RSVDsample} is carried out as a power iteration with subsequent QR factorizations to keep the sample orthonormal (see \cite{Halko2011LowRankProbabilistic}, Algorithm~4.4).
Step~\ref{alg:RSVDortho} then provides a basis of the sampled, relevant contributions to the range of $A$ in the orthonormal columns of the $m$-by-$\ell$ matrix $Q$. 
A rank-$\ell$ approximation of $A$ is now available by projection into that subspace: $A_\ell:=QQ^\dagger A$. 
Such an explicit construction is however not required. 
Instead, step~\ref{alg:RSVDtruncate} invokes a rank-$\bonddim$ \DSVD{} factorization $\tilde{U}\tilde{\Sigma}\tilde{V}^\dagger$ of the typically much smaller $\ell$-by-$n$ matrix $B:=Q^\dagger A$. 
Due to $A_\ell\approx A$, the $\bonddim$ largest singular values of $A$ are approximated in $\tilde{\Sigma}$. 
If required, approximate associated left- and right singular vectors of $A$ are given by $Q\tilde{U}$ and $\tilde{V}$, respectively. 
Both are exact isometries.

The average \RSVD{} compression error $\varepsilon_\text{\RSVD}:=\mathbb{E}\left(\|A-Q\tilde{U}\tilde{\Sigma}\tilde{V}^\dagger\|\right)$ depends on the spectrum of singular values, and can be made arbitrarily close to the minimal truncation error $\dtrunc$ in either Frobenius- or spectral norm: 
Following the analysis in \cite{Gu2015LowRankPIError}, a minimal oversampling of $\ell\ge\bonddim+2$ already guarantees
\begin{equation}
\varepsilon_\text{\RSVD} \le \sqrt{\dtrunc^2 + \mathcal{C}^2(n,\ell) \bonddim \sigma_{\ell-1}^2 \left(\sigma_{\ell-1} / \sigma_\bonddim\right)^{4q}}
\label{eq:error_rsvd}
\end{equation}
with $\mathcal{C}^2(n,\ell)$ in $\complexity(n\ell)$.
Note that $\dtrunc$ depends on the selected norm, unlike the additional terms introduced by the randomized approach.
While highest accuracy is expected for quickly decreasing singular values, a striking feature of \RSVD{} is that already small powers $q>0$ drive those contributions, which add to $\dtrunc$ in Eq.~\eqref{eq:error_rsvd}, to zero exponentially fast, even in cases of slowly decaying singular values. 
Furthermore, sufficient oversampling in $\ell$ makes the probability of a substantial deviation from the average error bound arbitrarily small \cite{Gu2015LowRankPIError}.

Throughout our benchmarks, we make the conservative choice $\ell=2\bonddim$, which is suitable to keep the \RSVD{}-error within a small factor of $\dtrunc$ even for $q=0$ \cite{Halko2011LowRankProbabilistic}.
In this configuration, \RSVD{} promises an asymptotic speedup over \DSVD{} in the order of the compression ratio
\begin{equation}
T_{\RSVDPREFIX}/T_{\DSVDPREFIX} \propto \bonddim/n \,,
\end{equation}
where $T_{\RSVDPREFIX}$ and $T_{\DSVDPREFIX}$ are the respective \RSVDPREFIX{}- and \DSVD{} run times on similar input $A$.
The proportionality is due to the lower \RSVD{} computational complexity, which is dominated by the matrix-matrix products  of $\complexity{\left(mn\ell(q+1)\right)}$ in sampling $Y$. 

The improved scaling of the \RSVD{} algorithm is complemented by its conceptual simplicity, which directly translates to a fast, stable and easily parallelizable implementation in terms of highly optimized linear algebra routines as provided by level-3 \textit{BLAS} and \textit{LAPACK} \cite{Anderson1999LAPACK_userguide}.
Various \RSVD{}-implementations are available, for instance in \textit{MATLAB}~\cite{Szlam2014RSVDinMATLAB}, in R~\cite{Erichson2016LowRankProbabilisticInR}, and via C-libraries such as \textit{RSVDPACK}~\cite{Voronin2015RSVDPACK} or the \textit{RRSVD-Package}~\cite{Tamascelli2015TEBDRSVD} which our benchmarks are based on.

\section{Benchmarks}\label{sec:BenchmarkSetup}

%
%
\begin{figure}
\includegraphics[width=1.\columnwidth]{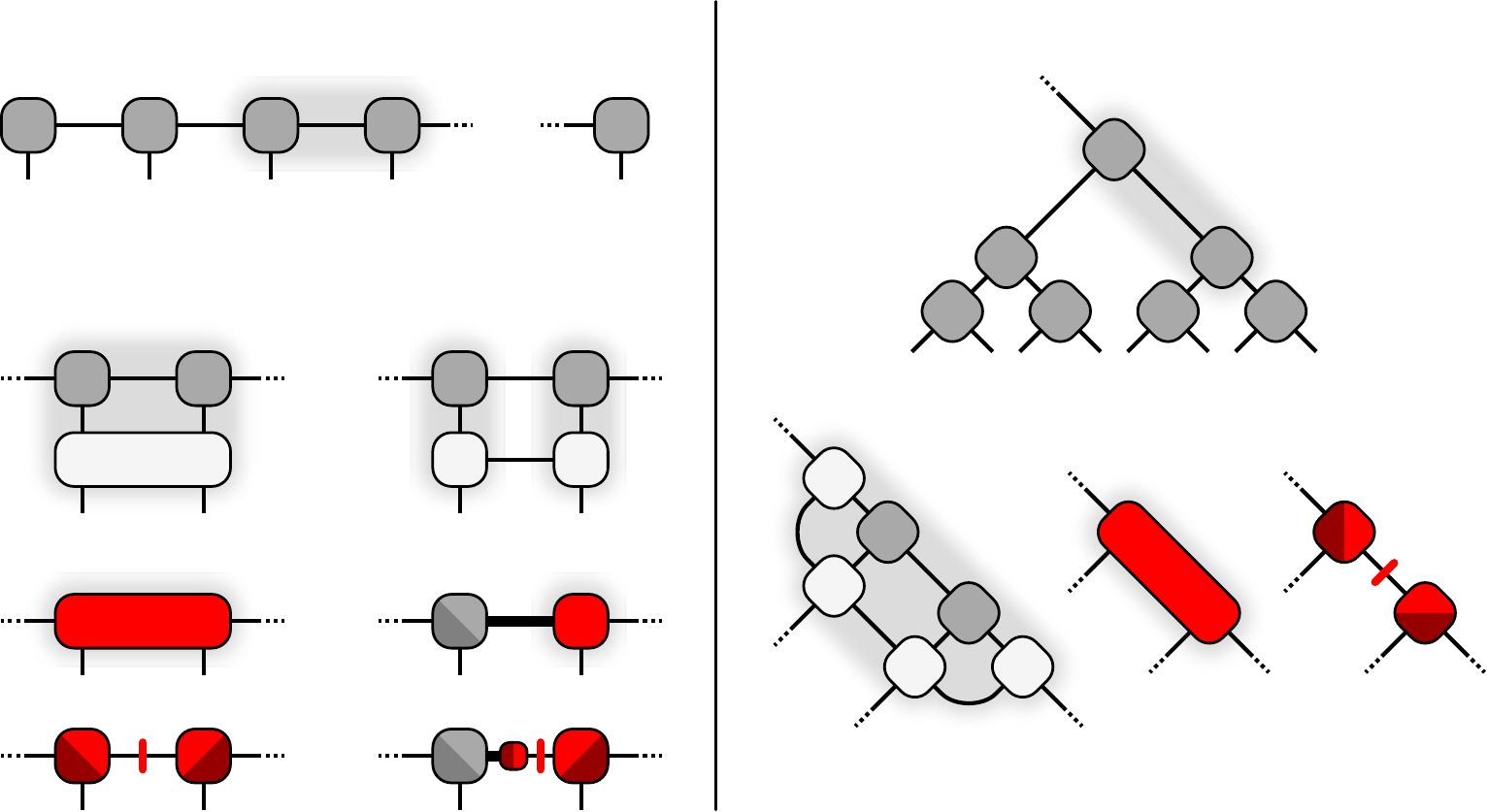}
\caption{ \label{fig:svd_update}
Compression steps in our benchmarked tensor networks of bond dimension $\bonddim$ and local dimension $\localdim$. 
The truncated SVD (highlighted in red) retains at most $\bonddim$ largest singular values (red bar) and produces isometric tensors (partially shaded).
Left: The TEBD algorithm updates the MPS (top) by absorbing nearest-neighbor evolution exponentials in the form of a full four-link tensor (B) or a sum over $K$ Kronecker products (P), resulting in different compression problems (bottom).
Right: In binary tree TNs (top), we minimize the energy of an effective Hamiltonian by jointly optimizing two adjacent tensors which are finally rank-$\bonddim$ factorized (bottom, left to right).
}
\end{figure}

We benchmark \RSVD{} against \DSVD{} performance, when employed in state-of-the-art TN algorithms.
Our focus lies on closed system ground states, and we compare both run time and precision of the relevant physical quantities in the outcomes.

We first outline the TN algorithms that drive our benchmark simulations and the role played by compression. 
Afterwards, we report model Hamiltonians and parameters. We close this section with a brief account on the numerical implementation.

\subsection{TN algorithms}\label{sub:TNalgorithms}

We employed TEBD imaginary time evolution on MPSs, and DMRG-style variational ground state search in the \bTTN{} \cite{Gerster2014TTN} with double-tensor optimization.
Both algorithms are well established techniques in ground state search of quantum lattice Hamiltonians.
They iteratively approximate those ground states in TNs of a selected maximal bond dimension $\bonddim$, defined over $d$-dimensional `physical' tensor indices that correspond to lattice sites (see Fig.~\ref{fig:svd_update}).
Specifically, both algorithms perform local update steps on adjacent tensors, which require a truncated SVD to recompress bond indices.
Note that it is the absence of loops (network cycles) in MPS and \bTTN{} geometries that makes truncated SVD an optimal protocol here, as it maintains maximum quantum fidelity between the states before and after the compression of a single bond \cite{White1992DMRG,White1993DMRG,Silvi2018LoopfreeTNMethods}
 
The two methods, however, rely on different local update steps:

In the TEBD algorithm, designed for time evolution with nearest-neighbor interactions, the update step consists of an application of a (real or imaginary) time-evolution exponential on two adjacent lattice sites. In standard TEBD, the exponential takes the form of a single four-index tensor or `block' (B) $u_\text{NN}$. 
It can also be given by a sum of Kronecker products (P) of single-site operators $\sum_{k=1}^K u_L^k \otimes u_R^k$, which can be more time- and memory efficient for $K<d$. 
Both strategies pose different compression problems (left side of Fig.~\ref{fig:svd_update}): 
The block-update contracts directly into a square matrix of dimension $\bonddim d$, while in the product-update we obtain a $(\bonddim K)$-by-$(\bonddim d)$ matrix instead. 
In both cases, the resulting matrix $A$ must be compressed into a rank-$\bonddim$ factorization with compression ratios $1/d$ and $1/\min(K,d)$, respectively. 
Consequently, we expect \RSVD{} to significantly speed up TEBD simulations on lattices with larger local dimensions $d$:
In terms of computational complexity, TEBD with typical bond dimension $\bonddim \ge d$ is dominated by the \DSVD{} compression step of $\complexity \left(\bonddim^3 d^3\right)$ in block- (B) and $\complexity \left(\bonddim^3 K^2 d\right)$ in product- (P) updates for $K\le d$.
The asymptotic \RSVD{} speedup can reduce this scaling to $\complexity \left(\bonddim^3 d^2\right)$ (for B), as demonstrated by \cite{Tamascelli2015TEBDRSVD}, and to $\complexity \left(\bonddim^3 K d\right)$ (for P), respectively -- which are typical costs exhibited by other operations within TEBD as well.

In the \bTTN{} setting, instead, the update step directly replaces two adjacent tensors with a matrix $A$ associated to the lowest eigenvector of an effective Hamiltonian. 
The matrix $A$ is at most a $\bonddim^2$-by-$\bonddim^2$ square matrix. 
On some lower levels of the tree geometry, smaller dimensions can be encountered, with $d^2$ at the physical indices on the bottom (right side of Fig.~\ref{fig:svd_update}). 
The majority of run time however is spent on the large update matrices, and these require a compression by a ratio $1/\bonddim$. 
A massive speedup of the compression step, in the order of the bond dimension, can thus be expected from employing \RSVD{} instead of \DSVD{}.

A feature of all simulations is that we explicitly target the symmetry-invariant ground states under certain global Abelian symmetries of the Hamiltonian. 
These grant us an inner block-structure in all tensors, which enhances efficiency and precision of the simulation \cite{Singh2010SymTN,Singh2011SymU1}. 
In the compression problem, we therefore encounter strictly block-diagonal matrices $A$, encoded in $\nsectors$ non-trivial blocks. 
The dimensions of these blocks correspond to degeneracies of symmetry sectors, and add up to the respective full dimensions of $A$. 
In all benchmarked situations, $\nsectors$ equals the (small) number of global symmetry sectors, and the optimal TN ground state approximations display more or less evenly sized block dimensions.
Since matrix factorizations can be done block-wise, all the actual matrix dimensions passed to the truncated SVD algorithm are thus roughly those of $A$ divided by $\nsectors$.
But as the truncation rank $\bonddim_s \approx \bonddim / \nsectors$ per block is similarly reduced, no change in the compression ratio and hence in the asymptotic speedup occurs.

Note that the truncation rank per block is usually not known a priori, as it depends on the number of large singular values $\sigma_j \ge \sigma_{\bonddim}$ therein.
This information is only directly available with \DSVD{}, where all singular values of all blocks are computed.
\RSVD{} on the other hand delivers just the requested number of singular values for each block, and some estimate of the appropriate truncation $\bonddim'_s \approx  \bonddim_s$ must be made beforehand. 
After \RSVD{}s are then performed in all sectors, we post-select the $\bonddim$ largest singular values and obtain the new optimal block dimensions $\bonddim_s$.
In our TEBD simulations we choose a block-wise truncation rank $\bonddim'_s = \bonddim/\nsectors + c$ with a small constant $c$ that allows for some variation in sector sizes (typically less than $5\%$). 
For \bTTN{}, we instead make the simplest maximal choice $\bonddim'_s = \bonddim$, which reduces the achievable speedup by a (small) factor $\nsectors$ but does not require any estimates.

\subsection{Models}

We simulated the quantum Ising model with ferromagnetic interaction in a tunable transverse field $\field$, on two different lattices: 
First, a 1D spin-$S$ chain of length $L$ with Hamiltonian
\begin{equation}
\label{eq:ising_chain}
\globaloptr{H}_\text{chain} = -\frac{1}{S^2} \sum_{j} X_j X_{j+1} + \frac{\field}{S} \sum_{j} Z_{j} \,,
\end{equation}
where $X$ and $Z$ are local spin operators (we set $\hbar=1$) and subscripts denote application sites. 
In general, in a computational spin-$Z$ eigenbasis $\left\{ \ket{m} \right\}$ of local dimension $\localdim=2S+1$ with integer or half-integer magnetic quantum numbers $m \in \left\{-S,-S+1,\dots,S\right\}$, we have
\begin{subequations}
	\begin{eqnarray}
	\braket{m'|Z|m} &=& m \times \delta_{m',m} \,,\\
	\braket{m'|X|m} &=& \sqrt{\left(S+1\right) \left(m+m'-1\right) - m m'}  \times \nonumber \\
	&&   \times \left(\delta_{m',m+1}+\delta_{m'+1,m}\right) / 2 \,.
	\end{eqnarray}
\end{subequations}
For $S=1/2$, $X$ and $Z$ reduce to standard Pauli matrices and the model is exactly solvable with quantum critical point at  $|h|=\fieldcrit=1$.
For $S\to\infty$, the transition point shifts with $\fieldcrit\to2$ \cite{Penson1984TransvIsingHighSpinCritical}. 
The TEBD is performed for $S>1/2$ in open boundary conditions with values of $\field$ in various distances to the critical points, which we estimated from finite-size scaling techniques \cite{Fisher1972FSScaling}.
In our \bTTN{} benchmark we focus exclusively on $S=1/2, \field=\fieldcrit$ in periodic boundary conditions. 

The second benchmark is the simulation of a spin-$1/2$ two-dimensional (2D) square-lattice Ising model in cylindrical boundary conditions of length $L$ and circumference (or width) $W$.
With respective site-subscripts $i$ and $j$, the Hamiltonian reads
\begin{equation}
\label{eq:ising_cylinder}
\globaloptr{H}_\text{cyl} = -\sum_{i,j}X_{i,j} X_{i,j+1} -\sum_{i,j}X_{i,j} X_{i+1,j} +\field\sum_{i,j}Z_{i,j} \,.
\end{equation}
By summation over $i$, we map this Hamiltonian onto an open chain of length $L$ with local dimension $\localdim=2^{W}$. 
For reasonably small values $W$, the ground state can be approximated in a MPS and its critical behavior can be studied with DMRG \cite{Jongh1998Ising2DDMRG}.
We performed imaginary TEBD at various values of $\field$, including points in proximity of the critical field at around $\fieldcrit \approx 3.044$, as reported with high precision in Monte Carlo and TN studies on the square lattice \cite{Bloete2002IsingMC,Rizzi2010MERACritical}.

As is well known, in the thermodynamic limit, the one- and two-dimensional Ising models of Eqs.~\eqref{eq:ising_chain} and \eqref{eq:ising_cylinder} exhibit spontaneous ferromagnetic order for $|\field|<\fieldcrit$, which breaks down in the paramagnetic phase for $|\field|>\fieldcrit$. 
Both phases are gapped, however at $|\field| = \fieldcrit$, the systems become critical and gapless. 

In the case of 1D lattices, we know that the ground states of a short-ranged, gapped system obey area laws for the entanglement entropy, while this is not true for a critical, gapless system \cite{Audenaert2002EntanglementChain,Plenio2005AreaLaw,Eisert2010AreaLaw,Wolf2006AreaLawFermions}.
Since  squares of the singular values in loop-free TN compression steps correspond to reduced density eigenvalues of lattice bipartitions, singular values are directly linked to bipartite entanglement measures such as the von~Neumann entropy, and thus the error analysis Eq.~\eqref{eq:error_rsvd} of \RSVD{} is linked to the physical properties of the ground state.
For this reason we perform our benchmarks at various values of $\field$, including values in close proximity to $\fieldcrit$. 
We expect the latter to pose the most demanding situation for \RSVD{} due to a potentially slow decay of tail singular values \cite{Calabrese2008CriticalSvals1D}, which make greater amounts of computational resources necessary (via parameters $q,\ell$) to avoid larger errors in Eq.~\eqref{eq:error_rsvd}.
As a comment, we remark that the benchmarked MPS and \bTTN{} simulations are best suited for non-critical systems due to finite bond dimensions $\bonddim$ that limit correlations and entanglement.
However, the selected finite lattice sizes admit simulations at and around $\field=\fieldcrit$, as is typical in extrapolating critical properties via finite-size scaling techniques \cite{Fisher1972FSScaling, Cardy2012FSScaling}. 
Furthermore, \bTTN{} have capabilities beyond MPS in encoding quantum critical ground states \cite{Silvi2010bTTN}.   

Both Ising models in Eqs.~\eqref{eq:ising_chain} and \eqref{eq:ising_cylinder} exhibit a global parity symmetry because their Hamiltonians commute with $\bigotimes_{j=1}^L {P}_j$, being defined locally by $\braket{m'|P|m} = \left(-1\right)^{m+S} \times \delta_{m',m}$. 
Local basis states transform as $P\ket{m^\pm}=\pm\ket{m^\pm}$ and fall either in the even `$+$' or odd sector `$-$' of dimensions $d_+$, $d_- \approx d/2$ respectively.
Rotations in the cylindrical boundary conditions provide an additional Abelian $Z_{W}$ cyclic symmetry for \eqref{eq:ising_cylinder}. 
As a consequence, even and odd sectors further decompose into $W$ different angular momentum sectors.
As mentioned in Sec.~\ref{sub:TNalgorithms}, we encode these symmetries explicitly, which allows us to restrict the TN state representation to the ground state global invariant sector $s=0$, that is the even parity and rotationally invariant subspace.

\subsection{Implementation}

Here we report the detailed implementation of a fair run-time- and precision comparison between \DSVD{} and \RSVD{}, and discuss technical details of the benchmarks.

We performed complete runs of our TEBD and \bTTN{} benchmark algorithms by iterating double-tensor updates until the energy expectation value of the TN state stagnates within some threshold $\precision$. 
Each run was repeated for different field $\field$, maximal bond dimension $\bonddim$, lattice length $L$ and a selected spin $S$ or width $W$, either with \DSVD{} or \RSVD{} in the compression steps.

For the precision comparison, we extracted expectations of energy and magnetization order, correlation- and entanglement properties and singular values from the produced final states.
The magnetization order $M$ was measured from nonlocal correlations,
\begin{equation}
M=\sqrt{ \sum_{k\neq k'} \braket{X_{k} X_{k'}} / \mathcal{N} } \,,
\label{eq:magnetization}
\end{equation}
where $k$ goes over all lattice sites and $\mathcal{N}$ counts the number of expectations summed over.
The estimate for the correlation-length $\bar{\xi}$ was computed from expectations values of $X_{(k)}\equiv X_k$ in the chain and $X_{(k)}\equiv X_{i,j}$ in the cylinder as follows:
\begin{equation}
\bar{\xi} = \sqrt{ \sum_{r>1} (r-1)^2  C_r / \sum_{r>1} {C_r} } \,.
\label{eq:corrlen}
\end{equation}
Here, $C_r$ denotes the bulk average over MPS sites $j$ of  $\braket{X_{(i,)j},X_{(i,)j+r}}$. 
The additional site index $i$ appears only in the two-dimensional model and is averaged over as well to extract only the `horizontal' correlation length subject to compression through the MPS bondlinks.
Note that $\bar{\xi}$ tends to underestimate the actual correlation-length and saturates below $L/\sqrt{6}$ if it becomes large compared to the system size.
Furthermore, profiles of the von~Neumann entropy $S_N(j)=-\sum_k \lambda^2_k \log(\lambda^2_k)$ have been obtained from the compressed singular values at MPS bonds $j=1,\dots,L-1$. 

We also profiled the individual run times spent in the truncated SVDs of compression steps, $T_{\DSVDPREFIX}$ and $T_{\RSVDPREFIX}$, and the time spent in all remaining parts of the algorithm,  $\overline{T}_{\DSVDPREFIX}$ and $\overline{T}_{\RSVDPREFIX}$ for \DSVD{} and \RSVD{} runs, respectively.
All these run times have been divided by the number of iterations performed in the simulation. 
However, we have found no substantial differences in the number of update steps performed with \DSVD{} and \RSVD{}, as reported in Sec.~\ref{sec:Results}.
We therefore obtain the average speedup in compression due to \RSVD{} from
\begin{equation}
\speedup := f \cdot T_{\DSVDPREFIX} / T_{\RSVDPREFIX} \,,
\label{eq:speedup}
\end{equation}
where $f:=\overline{T}_{\RSVDPREFIX} / \overline{T}_{\DSVDPREFIX}$ is the ratio of run times spent outside compression.
Since our benchmarks have been performed on shared cluster nodes, we introduced the factor $f$ to equalize the effect of the computational environment on the bare compression times.
Thus, simulation runs that were slowed done by other computations on a cluster node can be fairly compared to faster executed simulation runs. 

The complete simulation protocol for TEBD was as follows:
Starting from a product state with randomized tensors of bond dimension one, the algorithm is run in imaginary time with some sufficiently large initial time step $\timestep$ in the local imaginary time-evolution exponential. 
After a few first iterations out of typically many hundred, the bond dimension saturates the allowed maximum, and we can safely assume $\bonddim$ to be the typical compression rank. 
The simulation stops when convergence of the energy is detected as follows: 
Throughout the simulation, the change of the expectation value of the energy is monitored in regular intervals. 
Whenever this change drops below the targeted precision threshold $\precision$, the simulation time step $\timestep$ is subsequently reduced by a constant factor. 
Convergence is declared when the total energy decrement between two time-step reductions falls below $\precision$, too. 
With smaller choices of $\precision$, better approximations of the final MPS to the actual ground state of the system can be expected within the bond dimension $\bonddim$, but at the cost of increased number of iterations and run time.

The \bTTN{} ground state search employs randomized initial states remaining at maximal bond dimension throughout the entire simulation. 
The same initial states were used in comparative \DSVD{} and \RSVD{} runs. 
The algorithm then performs sequences of double-tensor updates on adjacent tensors, until the difference in energy expectation between subsequent sweeps falls below machine precision.

All simulations were carried out in double precision arithmetic with complex numbers, except for the imaginary TEBD on the spin-$1/2$ chain which we benchmarked in a TN representation with real elements, a common choice to enhance efficiency under time-reversal invariance.
Linear algebra computations (\textit{BLAS}, \textit{LAPACK}) where performed with the `Intel Math Kernel Library' (MKL) in versions 11.x.
Our fully truncated \DSVD{} implementation is based on the \textit{LAPACK} \textit{D/ZGESDD} divide-and-conquer algorithm.
For \RSVD{}, we employed the fixed-rank implementation from the \textit{RRSVD-Package} \cite{Tamascelli2015TEBDRSVD} with parameters $q=4, \ell=2\chi$ (see Sec.~\ref{sub:RSVDAlgorithm}) for any targeted truncation rank $\bonddim$.
This implementation employs \textit{LAPACK} \textit{D/ZGESVD} for the final factorization in step \ref{alg:RSVDtruncate} of the \RSVD{} algorithm.
All simulations were executed with single-threaded compression step on 16-way Intel Xeon E5--2670 (2.6 GHZ) compute nodes.


\section{Results}\label{sec:Results}

%
%
\begin{figure}
\includegraphics[width=1\columnwidth]{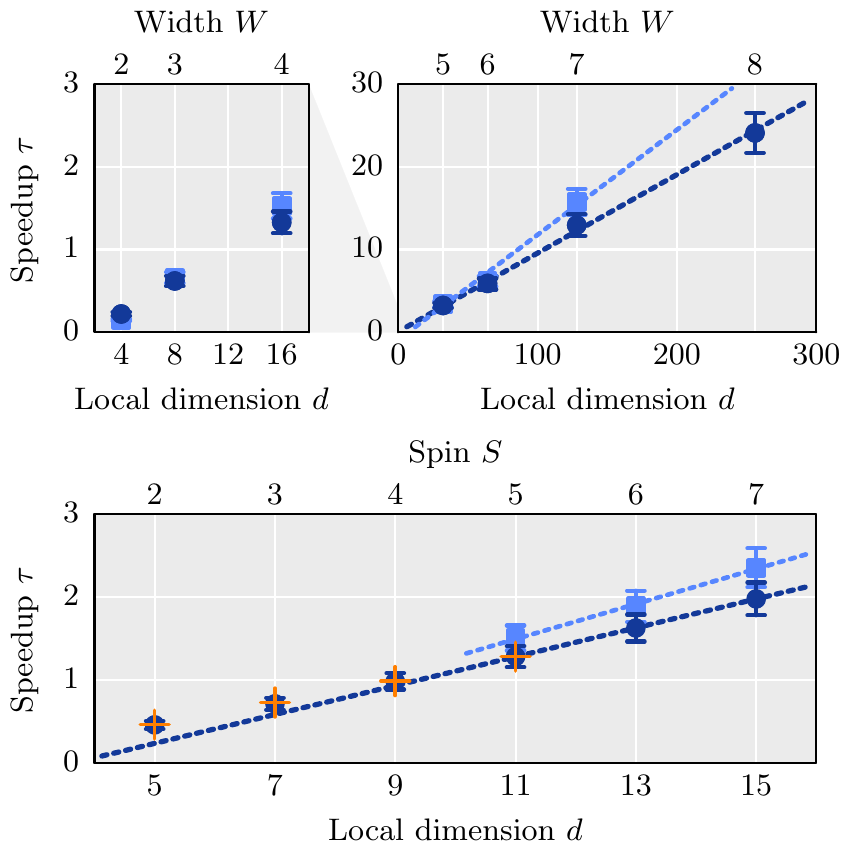}
\caption{ 
\label{fig:speedup_local}
Speedup in compression step due to \RSVD{} in TEBD simulations of increasing local dimensions. 
Top: 2D Ising model ($L=30$) as a function of the widths $W$. 
Bottom: 1D Ising model ($L=100$) as a function of local spin $S$.
Dark- and light blue points represent data at bond dimensions $\bonddim=100$ and $\bonddim=150$, respectively, both from `block'-update and fixed convergence criteria (1D: $\precision=10^{-13}$, 2D: $\precision=10^{-8}$ except $W=8$ was stopped before convergence). 
Error bars indicate a $10\%$ error estimate in speedups.
Orange crosses show speedup in higher precision target $\precision=10^{-14}$, $\bonddim=100$.
Dashed lines are linear fits for $\localdim>10$ with slopes $\approx 0.10,0.13$ (2D) and $0.18,0.21$ (1D) for $\bonddim=100,150$ respectively.
}
\end{figure}

%
%
\begin{figure}
\includegraphics[width=1\columnwidth]{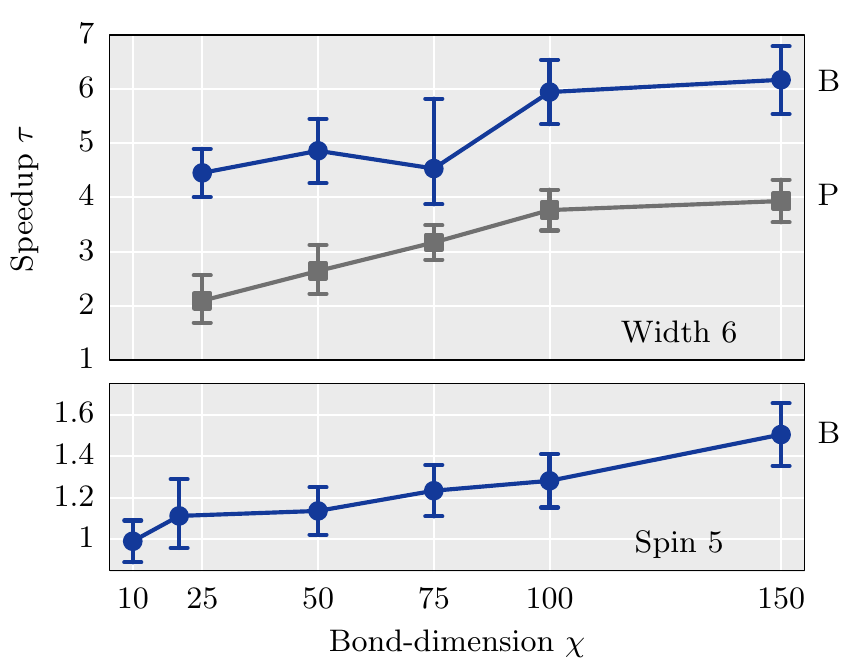}
\caption{ \label{fig:speedup_bond}
Dependency of \RSVD{}-speedup on the bond dimension in TEBD simulations. 
Top: 2D Ising model ($L=30$, $\precision=10^{-8}$) at width $W=6$. 
Blue and gray: `block' (B) and `product' (P) updates, respectively.
Bottom: 1D Ising model ($L=100$, $\precision=10^{-13}$) at spin $S=5$. 
}
\end{figure}

%
%
\begin{figure}
\includegraphics[width=1\columnwidth]{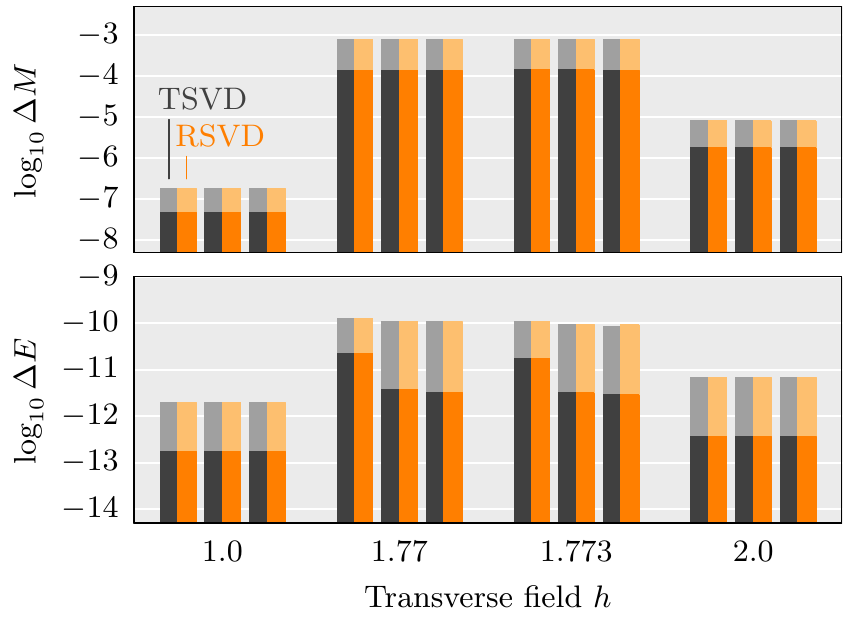}
\caption{ \label{fig:precision}
Relative error in 1D TEBD final state energy $\Delta E$ (top) and magnetization $\Delta M$ (bottom), compared to extrapolated ground state values $E_\text{best}$ and $M_\text{best}$ for $S=5, L=100$ and transverse fields in the ferro- ($\field=1.0$) and paramagnetic ($\field=2.0$) phases as well as close to the critical point.
Black and orange bars correspond to \DSVDPREFIX{}- and \RSVD{} results, respectively, at convergence thresholds $\precision=10^{-11}$ and $10^{-13}$ (light shaded).
Each group of three bars shows the error with increasing bond dimensions $\bonddim=50,75,100$ (left to right).
}
\end{figure}

%
%
\begin{figure}
\includegraphics[width=1\columnwidth]{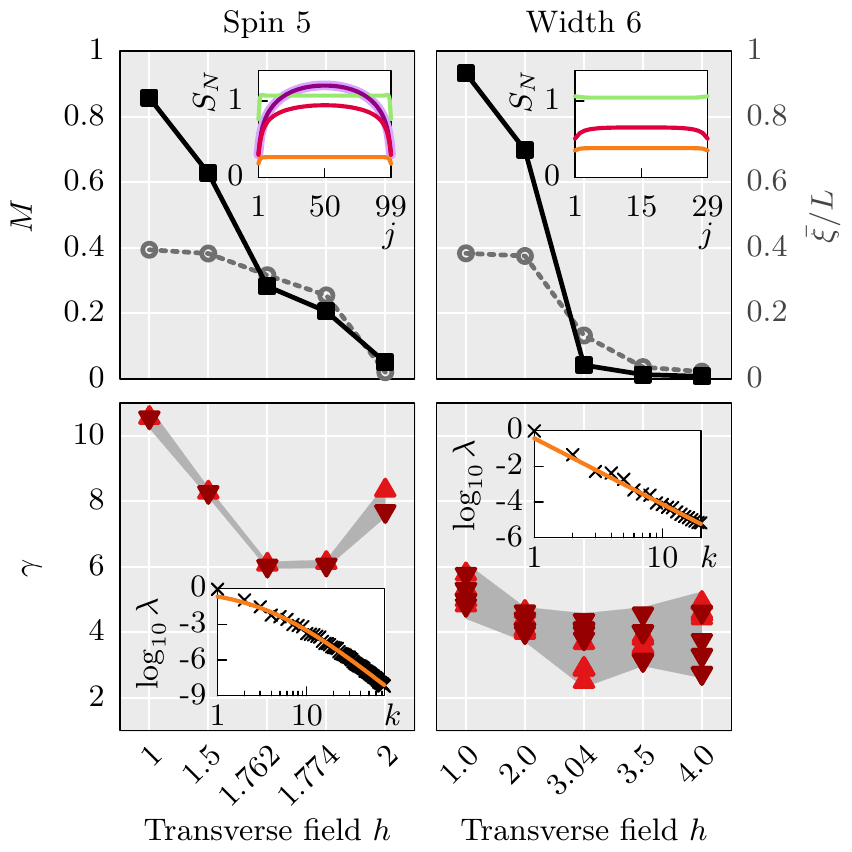}
\caption{ \label{fig:critical_props}
Correlation properties and singular values in TEBD simulated ground states of 1D (left, $S=5,L=100$) and 2D Ising models (right, $W=6,L=30$) at various transverse fields $h$.
Top panels: Magnetization $M$ (black) and estimates for the correlation length $\bar{\xi}/L$ (gray dashed). 
Errors are smaller than point sizes.
Top insets: Von~Neumann Entropies $S_N(j)$ from singular values on MPS bonds. 
1D, left: $h=1.5$ (green), $1.762$ (purple with light-purple fit $S^C_N(j)$, see text), $1.774$ (red), 2.0 (orange). 
2D, right: $h=2.0$ (green), $3.04$ (red), $3.5$ (orange).
Bottom insets: Singular values (black) at a central bond in the invariant sector for near-critical fields $h=1.7735$ in 1D (left) and $h=3.04$ in 2D (right) and a polynomial fit $\sigma_k\approx (C_1 k + C_2)^{-\gamma}$ (orange, $C_2=0$ in 2D).
Bottom panels: Decay exponents $\gamma$ of singular values. 
Up/downwards pointing triangles indicate even/odd sectors, respectively.
Shaded area encloses fit errors.
}
\end{figure}

%
%
\begin{figure}
\includegraphics[width=1\columnwidth]{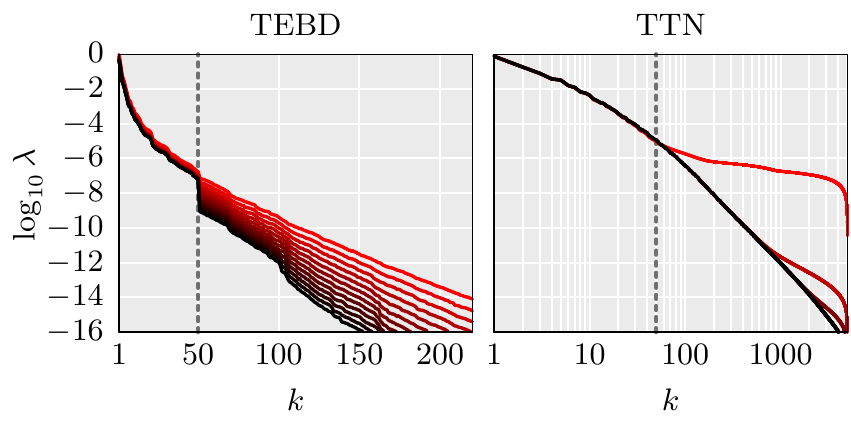}
\caption{ \label{fig:svals_decay}
Singular values $\lambda_k$ monitored over algorithm runtime in 1D Ising models at, or close to, the critical field. 
Both panels show values in the invariant sector at a central lattice bipartition with $\bonddim=100$.
The dashed lines indicate the truncation at $\bonddim_{0}=50$.
Left: Imaginary TEBD ($S=5$, $L=100$, $\field=1.7735$, $\precision=10^{-13}$, block-update `B').
The time step $\timestep$ was subsequently reduced to $\timestep_i=0.4\times 0.7^i$ for $i=0,\dots,11$ (red to black).
Right: \bTTN{} ground state search ($S=1/2$, $L=128$, $\field=1.0$), after $i=0,\dots,4$ network updates (red to black).
}
\end{figure}

We first report the speedups obtained from upgrading compression steps from \DSVD{} to \RSVD{}. 
We then present evidence that no loss of precision occurs due to \RSVD{}. 
Finally we present selected ground state properties and spectra of singular values that we encountered in our benchmarks. 

All the following speedups have been obtained from independent simulations according to Eq.~\eqref{eq:speedup} with an estimated uncertainty of at most $\Delta \speedup \approx 10\,\%$.
Equal numbers of \RSVD{} and \DSVD{} compression steps were performed in all \bTTN{} simulations. 
Some imaginary TEBD runs converged in less iterations with either \RSVD{} or \DSVD{}, but those fluctuations were negligible compared to $\Delta \speedup$.

Speedups up to $\tau\approx24$ have been reached in TEBD simulations of increasing local dimensions, as shown in Fig.~\ref{fig:speedup_local} for the one- and two-dimensional Ising models of Eqs.~\eqref{eq:ising_chain} and \eqref{eq:ising_cylinder}. 
We observe that \RSVD{} outperforms \DSVD{} for $d > 10$, with speedups directly proportional to $d$ as predicted by the asymptotic cost analysis in Sec.~\ref{sub:TNalgorithms}.
These speedups remain stable under different algorithm parameters, such as changes in convergence criteria (orange crosses in bottom panel of Fig.~\ref{fig:speedup_local}).
We also found no significant dependency on the transverse field $\field$:
Thus, all speedups are geometric means over five (2D) and ten (1D) different values of $\field$ in various distances from (including close proximity to) the critical point, and each speedup falls within the error bars.

In all cases however, the speedup tends to increase with the bond dimension, as shown in Fig.~\ref{fig:speedup_bond} for selected one- and two-dimensional TEBD simulation.
The latter suggests some saturation at high bond dimension.
Again, all speedups shown are geometric means over at least ten simulations at transverse fields $\field$ in various distances from $\fieldcrit$, which had no significant impact on the speedup, as can be seen from the error bars that always enclose minimal and maximal speedup. 

In support of our TEBD results, the \bTTN{} benchmarks demonstrates massive \RSVD{} speedups $\speedup$ when bond dimensions are scaled up:
For instance at $\bonddim=60$ we found $\speedup\approx6$, while $\bonddim=100$ already provided us with $\speedup\approx11$ on a spin-$1/2$ lattice of length $L=64$.

Next, we assess the accuracy of the final states delivered by our \DSVD{} and \RSVD{} benchmarks.
To this end, we compare the simulation errors in energy expectation value $E$ and non-local magnetization order parameter $M$ of Eq.~\eqref{eq:magnetization} for various simulation parameters such as $\field$, $\bonddim$ and precision target $\precision$.
The errors are computed from differences $\Delta E = (E - E_\text{best})/E_\text{best}$ and $\Delta M = \left|M - M_\text{best}\right|/M_\text{best}$ to high precision data $E_\text{best}$ and $M_\text{best}$, respectively.
In TEBD simulations, $E_\text{best}$ and $M_\text{best}$ have been extrapolated from bond dimensions and precisions up to $\bonddim=150$, $\precision=10^{-14}$ using \DSVD{}, with uncertainty smaller than all observed differences $\Delta E$ and $\Delta M$ (typically one or more orders of magnitude).
We found that both \DSVD{} and \RSVD{} produce comparable simulation errors in all benchmarks, as exemplified in Fig.~\ref{fig:precision} for TEBD simulations of the one-dimensional Ising model for $L=100$ and $S=5$. 
We found similar results for up to $L=400$ in various precision targets and bond dimensions $\bonddim \le 100$ in both para- and ferromagnetic phases as well as close to the critical point.
In two-dimensional TEBD simulations at $W=6$, $L=30$ and in the \bTTN{} benchmarks, \DSVD{}- and \RSVD{} results even matched within computational precision.

The range of physical properties covered by our benchmarks is demonstrated in Fig.~\ref{fig:critical_props}, where the upper panel shows the magnetization $M$ and the estimate for the correlation length $\bar{\xi}$ (see Eq.~\eqref{eq:corrlen}) in the final TEBD simulation states.
These results, taken from \DSVD{} runs of 1D and 2D Ising models for some of the benchmarked transverse fields $h$, display values of magnetic order and correlation lengths spanning the entire spectrum of possible outcomes.
Furthermore, the von~Neumann entropies $S_N(j)$ on the MPS bonds confirm area-laws in both ordered and unordered phases as well as typical corrections near the 1D critical point, which are well described by a fit to $S^C_N(j) = a + c/6 \log\{ L/\pi \sin( \pi j/L ) \}$ with some constants $a,c$ \cite{Calabrese2009EntanglementEntropyReview}.
The corresponding singular values are detailed in the bottom panel of Fig.~\ref{fig:critical_props}.
Within the bond dimensions $\bonddim_s$ of individual symmetry sectors, they are well fitted by power-law decays $\sigma_k\approx (C_1 k + C_2)^{-\gamma}$ with fit constants $C_1$, $C_2$ and decay exponents $\gamma$ ranging from $-2$ to $-11$.

This decay of singular values, which relates physical properties to RSVD performace (as discussed in Sec.~\ref{sec:BenchmarkSetup}) is further analyzed in Fig.~\ref{fig:svals_decay} where we present complete spectra of singular values from the local compression problems $A$, including the truncated tail of small singular values, for a central bond and critical transverse field. 
In both TEBD (left panel) and \bTTN{} (right panel) simulations, the spectrum of singular values $\lambda_k$ can be separated into two parts:
For $k\le\bonddim_s$, the spectrum appears to undergo only minor changes throughout the algorithm run time, and is well described by the actual decay in the final (ground) state (see Fig.~\ref{fig:critical_props} for TEBD) over the majority of the run time.
For $k>\bonddim_s$, on the other hand, we observe a tail spectrum that does not necessarily follow the characteristics expected from the actual ground state (i.e. $\bonddim\to\infty$).
Namely, it changes significantly over the algorithm run time, and exhibits the fastest decay in the final iteration(s) of the algorithm: 
In case of TEBD, the tail can be seen to be bounded by a rapid polynomial decay, well separated from the retained singular values as it finally becomes proportional to a very small evolution time step $\timestep$.
In \bTTN{}, compression starts from a rather flat tail spectrum, that quickly approaches an exponential decay.
This demonstrates that the compression problem within the TN approximation becomes increasingly well conditioned for \RSVD{}, even close to the phase transition, as the algorithm converges closer to the ground state. 
This allows \RSVD{} to deliver higher precision (cf. Eq.~\eqref{eq:error_rsvd}, due to oversampling) with higher reliability right in the final stages of the algorithms when most needed.


\section{Discussion and Outlook}\label{sec:Discussion}

We provided evidence for substantially accelerated compression of tensor networks in all benchmarked algorithms by simply replacing the full truncated \DSVD{} with the \RSVD{} algorithm.
In particular, \RSVD{} outperformed \DSVD{} with the expected asymptotic speedup, that is inversely proportional to the compression ratio, when not more than $10\%$ of singular values were retained.

Remarkably enough, we attained those speedups without loss of precision in the simulated ground states: 
With \RSVD{} we reproduced local expectation values such as the energy, as well as long-range correlation- and entanglement properties, with differences to \DSVD{} simulations far smaller than the inherent ansatz errors due to a finite bond dimension or number of iterations performed. 

By benchmarking with encoded Abelian symmetries, we confirmed the \RSVD{} speedup in reduced bond- and local dimensions per sector.
Even though small matrix sizes can reduce speedups, \RSVD{} becomes increasingly useful with the typically large bond dimensions that are required for ground state approximation.

All results, moreover, hold up independently from the various physical scenarios, i.e.\ off- and at quantum critical points over a wide range of correlation lengths and respective spectra of the singular values.

Remarkably, the iterative nature shared by many TN algorithms has been observed to work in favor of \RSVD{} in that the truncated tail singular values decayed fast in the relevant final iterations, even close to phase transitions.

We expect the presented results to be robust and reproducible in a wide range of tensor network applications.
For instance, our choice of \RSVD{}-parameters ($q,\ell$) has been extremely conservative, as confirmed by the small differences to \DSVD{} in the outcomes, and can be fine tuned for much higher efficiency: 
Namely, by reducing $q$, \RSVD{} might outperform \DSVD{} for compression ratios as moderate as $20\%$ or less. 
With \RSVD{}, precision and efficiency of the compression can further be balanced dynamically, which promises significant reduction of run time in the earlier algorithm stages, as is already standard practice for instance in the eigensolver optimization steps in DMRG.
In this regard, it may prove specifically useful that \RSVD{} can also deliver a fixed error (instead of fixed rank) approximation: 
Parameters such as $\bonddim$, $\ell$ and possibly $q$ are then dynamically adjusted to deliver a compression within a given error bound \cite{Halko2011LowRankProbabilistic,Gu2015LowRankPIError}. 
Such dynamics might also provide an alternative route to fix the compressed sector sizes $\bonddim_s$ in presence of symmetric TN, even though good estimates (for instance based on previous iterations) plus added oversampling work well as demonstrated. 
Moreover, ongoing development of the \RSVD{} method itself may lead to further optimizations, such as modified power iteration schemes for faster convergence  \cite{Musco2015RandomizedKrylowSVD} or single view algorithms \cite{Tropp2016SingleView}.

With the benchmarked ground state simulations, it is clear that \RSVD{} is indeed not limited to open system real time dynamics with TEBD \cite{Tamascelli2015TEBDRSVD}, and we foresee a broad impact on DMRG and imaginary- or real time evolution codes that operate on ground states, including short-time quenches \cite{Eisert2006QuenchEntanglement} out of equilibrium via TEBD or the time-dependent variational principle \cite{Haegeman2011TDVP}.
This in turn could open new possibilities, for instance, in the numerical study of the Kibble-Zurek mechanism \cite{Kibble1976Cosmic,Zurek1985Cosmological}.
More generally, \RSVD{} has great potential in all scenarios that make extensive use of truncated SVD with small compression ratios.
Those arise naturally in TN algorithms that operate on potentially large tensors and in various double-tensor update strategies, that are regularly employed in DMRG and time evolution codes when Abelian- or non-Abelian symmetries are encoded, and to avoid meta-stabilities that hinder convergence \cite{White2005SingleCenterDMRG,Hubig2015DMRGSubspaceExpansion}.
Such scenarios include, for example, applications to higher dimensions, lattice models with large local dimensions or applications of TNs in quantum chemistry.


\begin{acknowledgments}
We thank Matthias Gerster for discussions and sharing his \bTTN{} code.
The authors gratefully acknowledge support from the Carl-Zeiss-Stiftung via Nachwuchsf\"orderprogramm, the state of Baden-W\"urttemberg through bwHPC, the Italian HPC facility CINECA through the TEDDI project and the German Research Foundation (DFG) through the TWITTER grants. 
S.M. gratefully acknowledges the support of the DFG via a Heisenberg fellowship.
This work was supported by the ERC Synergy grant BioQ.
\end{acknowledgments}

\vspace{1em}

\begin{thebibliography}{75}%
\makeatletter
\providecommand \@ifxundefined [1]{%
 \@ifx{#1\undefined}
}%
\providecommand \@ifnum [1]{%
 \ifnum #1\expandafter \@firstoftwo
 \else \expandafter \@secondoftwo
 \fi
}%
\providecommand \@ifx [1]{%
 \ifx #1\expandafter \@firstoftwo
 \else \expandafter \@secondoftwo
 \fi
}%
\providecommand \natexlab [1]{#1}%
\providecommand \enquote  [1]{``#1''}%
\providecommand \bibnamefont  [1]{#1}%
\providecommand \bibfnamefont [1]{#1}%
\providecommand \citenamefont [1]{#1}%
\providecommand \href@noop [0]{\@secondoftwo}%
\providecommand \href [0]{\begingroup \@sanitize@url \@href}%
\providecommand \@href[1]{\@@startlink{#1}\@@href}%
\providecommand \@@href[1]{\endgroup#1\@@endlink}%
\providecommand \@sanitize@url [0]{\catcode `\\12\catcode `\$12\catcode
  `\&12\catcode `\#12\catcode `\^12\catcode `\_12\catcode `\%12\relax}%
\providecommand \@@startlink[1]{}%
\providecommand \@@endlink[0]{}%
\providecommand \url  [0]{\begingroup\@sanitize@url \@url }%
\providecommand \@url [1]{\endgroup\@href {#1}{\urlprefix }}%
\providecommand \urlprefix  [0]{URL }%
\providecommand \Eprint [0]{\href }%
\providecommand \doibase [0]{http://dx.doi.org/}%
\providecommand \selectlanguage [0]{\@gobble}%
\providecommand \bibinfo  [0]{\@secondoftwo}%
\providecommand \bibfield  [0]{\@secondoftwo}%
\providecommand \translation [1]{[#1]}%
\providecommand \BibitemOpen [0]{}%
\providecommand \bibitemStop [0]{}%
\providecommand \bibitemNoStop [0]{.\EOS\space}%
\providecommand \EOS [0]{\spacefactor3000\relax}%
\providecommand \BibitemShut  [1]{\csname bibitem#1\endcsname}%
\let\auto@bib@innerbib\@empty
\bibitem [{\citenamefont
  {Schollw{\"o}ck}(2011)}]{Schollwoeck2011DMRG_MPS_review}%
  \BibitemOpen
  \bibfield  {author} {\bibinfo {author} {\bibfnamefont {U.}~\bibnamefont
  {Schollw{\"o}ck}},\ }\href {\doibase 10.1016/j.aop.2010.09.012} {\bibfield
  {journal} {\bibinfo  {journal} {Annals of Physics}\ }\textbf {\bibinfo
  {volume} {326}},\ \bibinfo {pages} {96} (\bibinfo {year} {2011})}\BibitemShut
  {NoStop}%
\bibitem [{\citenamefont {Or{\'{u}}s}(2014)}]{Orus2014TN_review}%
  \BibitemOpen
  \bibfield  {author} {\bibinfo {author} {\bibfnamefont {R.}~\bibnamefont
  {Or{\'{u}}s}},\ }\href {\doibase 10.1016/j.aop.2014.06.013} {\bibfield
  {journal} {\bibinfo  {journal} {Annals of Physics}\ }\textbf {\bibinfo
  {volume} {349}},\ \bibinfo {pages} {117} (\bibinfo {year}
  {2014})}\BibitemShut {NoStop}%
\bibitem [{\citenamefont {White}(1992)}]{White1992DMRG}%
  \BibitemOpen
  \bibfield  {author} {\bibinfo {author} {\bibfnamefont {S.~R.}\ \bibnamefont
  {White}},\ }\href {\doibase 10.1103/physrevlett.69.2863} {\bibfield
  {journal} {\bibinfo  {journal} {Physical Review Letters}\ }\textbf {\bibinfo
  {volume} {69}},\ \bibinfo {pages} {2863} (\bibinfo {year}
  {1992})}\BibitemShut {NoStop}%
\bibitem [{\citenamefont {White}(1993)}]{White1993DMRG}%
  \BibitemOpen
  \bibfield  {author} {\bibinfo {author} {\bibfnamefont {S.~R.}\ \bibnamefont
  {White}},\ }\href {\doibase 10.1103/physrevb.48.10345} {\bibfield  {journal}
  {\bibinfo  {journal} {Physical Review B}\ }\textbf {\bibinfo {volume} {48}},\
  \bibinfo {pages} {10345} (\bibinfo {year} {1993})}\BibitemShut {NoStop}%
\bibitem [{\citenamefont {Rommer}\ and\ \citenamefont
  {{\"{O}}stlund}(1997)}]{Rommer1997MPS}%
  \BibitemOpen
  \bibfield  {author} {\bibinfo {author} {\bibfnamefont {S.}~\bibnamefont
  {Rommer}}\ and\ \bibinfo {author} {\bibfnamefont {S.}~\bibnamefont
  {{\"{O}}stlund}},\ }\href {\doibase 10.1103/physrevb.55.2164} {\bibfield
  {journal} {\bibinfo  {journal} {Physical Review B}\ }\textbf {\bibinfo
  {volume} {55}},\ \bibinfo {pages} {2164} (\bibinfo {year}
  {1997})}\BibitemShut {NoStop}%
\bibitem [{\citenamefont {Eisert}(2013)}]{Eisert2013TNReview_Entanglement}%
  \BibitemOpen
  \bibfield  {author} {\bibinfo {author} {\bibfnamefont {J.}~\bibnamefont
  {Eisert}},\ }\href@noop {} {\enquote {\bibinfo {title} {Entanglement and
  tensor network states},}\ } (\bibinfo {year} {2013}),\ \Eprint
  {http://arxiv.org/abs/arXiv:1308.3318v2} {arXiv:1308.3318v2} \BibitemShut
  {NoStop}%
\bibitem [{\citenamefont {Audenaert}\ \emph {et~al.}(2002)\citenamefont
  {Audenaert}, \citenamefont {Eisert}, \citenamefont {Plenio},\ and\
  \citenamefont {Werner}}]{Audenaert2002EntanglementChain}%
  \BibitemOpen
  \bibfield  {author} {\bibinfo {author} {\bibfnamefont {K.}~\bibnamefont
  {Audenaert}}, \bibinfo {author} {\bibfnamefont {J.}~\bibnamefont {Eisert}},
  \bibinfo {author} {\bibfnamefont {M.~B.}\ \bibnamefont {Plenio}}, \ and\
  \bibinfo {author} {\bibfnamefont {R.~F.}\ \bibnamefont {Werner}},\ }\href
  {\doibase 10.1103/PhysRevA.66.042327} {\bibfield  {journal} {\bibinfo
  {journal} {Phys. Rev. A}\ }\textbf {\bibinfo {volume} {66}},\ \bibinfo
  {pages} {042327} (\bibinfo {year} {2002})}\BibitemShut {NoStop}%
\bibitem [{\citenamefont {Plenio}\ \emph {et~al.}(2005)\citenamefont {Plenio},
  \citenamefont {Eisert}, \citenamefont {Drei\ss{}ig},\ and\ \citenamefont
  {Cramer}}]{Plenio2005AreaLaw}%
  \BibitemOpen
  \bibfield  {author} {\bibinfo {author} {\bibfnamefont {M.~B.}\ \bibnamefont
  {Plenio}}, \bibinfo {author} {\bibfnamefont {J.}~\bibnamefont {Eisert}},
  \bibinfo {author} {\bibfnamefont {J.}~\bibnamefont {Drei\ss{}ig}}, \ and\
  \bibinfo {author} {\bibfnamefont {M.}~\bibnamefont {Cramer}},\ }\href
  {\doibase 10.1103/PhysRevLett.94.060503} {\bibfield  {journal} {\bibinfo
  {journal} {Phys. Rev. Lett.}\ }\textbf {\bibinfo {volume} {94}},\ \bibinfo
  {pages} {060503} (\bibinfo {year} {2005})}\BibitemShut {NoStop}%
\bibitem [{\citenamefont {Eisert}\ \emph {et~al.}(2010)\citenamefont {Eisert},
  \citenamefont {Cramer},\ and\ \citenamefont {Plenio}}]{Eisert2010AreaLaw}%
  \BibitemOpen
  \bibfield  {author} {\bibinfo {author} {\bibfnamefont {J.}~\bibnamefont
  {Eisert}}, \bibinfo {author} {\bibfnamefont {M.}~\bibnamefont {Cramer}}, \
  and\ \bibinfo {author} {\bibfnamefont {M.~B.}\ \bibnamefont {Plenio}},\
  }\href {\doibase 10.1103/revmodphys.82.277} {\bibfield  {journal} {\bibinfo
  {journal} {Reviews of Modern Physics}\ }\textbf {\bibinfo {volume} {82}},\
  \bibinfo {pages} {277} (\bibinfo {year} {2010})}\BibitemShut {NoStop}%
\bibitem [{\citenamefont {Wolf}(2006)}]{Wolf2006AreaLawFermions}%
  \BibitemOpen
  \bibfield  {author} {\bibinfo {author} {\bibfnamefont {M.~M.}\ \bibnamefont
  {Wolf}},\ }\href {\doibase 10.1103/PhysRevLett.96.010404} {\bibfield
  {journal} {\bibinfo  {journal} {Phys. Rev. Lett.}\ }\textbf {\bibinfo
  {volume} {96}},\ \bibinfo {pages} {010404} (\bibinfo {year}
  {2006})}\BibitemShut {NoStop}%
\bibitem [{\citenamefont {Verstraete}\ \emph {et~al.}(2006)\citenamefont
  {Verstraete}, \citenamefont {Wolf}, \citenamefont {Perez-Garcia},\ and\
  \citenamefont {Cirac}}]{Verstraete2006PEPS}%
  \BibitemOpen
  \bibfield  {author} {\bibinfo {author} {\bibfnamefont {F.}~\bibnamefont
  {Verstraete}}, \bibinfo {author} {\bibfnamefont {M.~M.}\ \bibnamefont
  {Wolf}}, \bibinfo {author} {\bibfnamefont {D.}~\bibnamefont {Perez-Garcia}},
  \ and\ \bibinfo {author} {\bibfnamefont {J.~I.}\ \bibnamefont {Cirac}},\
  }\href {\doibase 10.1103/PhysRevLett.96.220601} {\bibfield  {journal}
  {\bibinfo  {journal} {Phys. Rev. Lett.}\ }\textbf {\bibinfo {volume} {96}},\
  \bibinfo {pages} {220601} (\bibinfo {year} {2006})}\BibitemShut {NoStop}%
\bibitem [{\citenamefont {Tagliacozzo}\ \emph {et~al.}(2009)\citenamefont
  {Tagliacozzo}, \citenamefont {Evenbly},\ and\ \citenamefont
  {Vidal}}]{Tagliacozzo2009TTN2D}%
  \BibitemOpen
  \bibfield  {author} {\bibinfo {author} {\bibfnamefont {L.}~\bibnamefont
  {Tagliacozzo}}, \bibinfo {author} {\bibfnamefont {G.}~\bibnamefont
  {Evenbly}}, \ and\ \bibinfo {author} {\bibfnamefont {G.}~\bibnamefont
  {Vidal}},\ }\href {\doibase 10.1103/PhysRevB.80.235127} {\bibfield  {journal}
  {\bibinfo  {journal} {Phys. Rev. B}\ }\textbf {\bibinfo {volume} {80}},\
  \bibinfo {pages} {235127} (\bibinfo {year} {2009})}\BibitemShut {NoStop}%
\bibitem [{\citenamefont {Gerster}\ \emph {et~al.}(2017)\citenamefont
  {Gerster}, \citenamefont {Rizzi}, \citenamefont {Silvi}, \citenamefont
  {Dalmonte},\ and\ \citenamefont {Montangero}}]{Gerster2017bTTNHofstadter}%
  \BibitemOpen
  \bibfield  {author} {\bibinfo {author} {\bibfnamefont {M.}~\bibnamefont
  {Gerster}}, \bibinfo {author} {\bibfnamefont {M.}~\bibnamefont {Rizzi}},
  \bibinfo {author} {\bibfnamefont {P.}~\bibnamefont {Silvi}}, \bibinfo
  {author} {\bibfnamefont {M.}~\bibnamefont {Dalmonte}}, \ and\ \bibinfo
  {author} {\bibfnamefont {S.}~\bibnamefont {Montangero}},\ }\href@noop {} {\
  (\bibinfo {year} {2017})},\ \Eprint {http://arxiv.org/abs/arXiv:1308.3318v2}
  {arXiv:1308.3318v2} \BibitemShut {NoStop}%
\bibitem [{\citenamefont {Vidal}(2007)}]{Vidal2007MERA}%
  \BibitemOpen
  \bibfield  {author} {\bibinfo {author} {\bibfnamefont {G.}~\bibnamefont
  {Vidal}},\ }\href {\doibase 10.1103/physrevlett.99.220405} {\bibfield
  {journal} {\bibinfo  {journal} {Physical Review Letters}\ }\textbf {\bibinfo
  {volume} {99}},\ \bibinfo {pages} {220405} (\bibinfo {year}
  {2007})}\BibitemShut {NoStop}%
\bibitem [{\citenamefont {Vidal}(2008)}]{Vidal2008MERA}%
  \BibitemOpen
  \bibfield  {author} {\bibinfo {author} {\bibfnamefont {G.}~\bibnamefont
  {Vidal}},\ }\href {\doibase 10.1103/PhysRevLett.101.110501} {\bibfield
  {journal} {\bibinfo  {journal} {Phys. Rev. Lett.}\ }\textbf {\bibinfo
  {volume} {101}},\ \bibinfo {pages} {110501} (\bibinfo {year}
  {2008})}\BibitemShut {NoStop}%
\bibitem [{\citenamefont {Silvi}\ \emph {et~al.}(2010)\citenamefont {Silvi},
  \citenamefont {Giovannetti}, \citenamefont {Montangero}, \citenamefont
  {Rizzi}, \citenamefont {Cirac},\ and\ \citenamefont {Fazio}}]{Silvi2010bTTN}%
  \BibitemOpen
  \bibfield  {author} {\bibinfo {author} {\bibfnamefont {P.}~\bibnamefont
  {Silvi}}, \bibinfo {author} {\bibfnamefont {V.}~\bibnamefont {Giovannetti}},
  \bibinfo {author} {\bibfnamefont {S.}~\bibnamefont {Montangero}}, \bibinfo
  {author} {\bibfnamefont {M.}~\bibnamefont {Rizzi}}, \bibinfo {author}
  {\bibfnamefont {J.~I.}\ \bibnamefont {Cirac}}, \ and\ \bibinfo {author}
  {\bibfnamefont {R.}~\bibnamefont {Fazio}},\ }\href {\doibase
  10.1103/physreva.81.062335} {\bibfield  {journal} {\bibinfo  {journal}
  {Physical Review A}\ }\textbf {\bibinfo {volume} {81}},\ \bibinfo {pages}
  {062335} (\bibinfo {year} {2010})}\BibitemShut {NoStop}%
\bibitem [{\citenamefont {Or\'us}\ \emph {et~al.}(2014)\citenamefont {Or\'us},
  \citenamefont {Wei}, \citenamefont {Buerschaper},\ and\ \citenamefont
  {Garc\'{\i}a-Saez}}]{Orus2014MultipartiteTN}%
  \BibitemOpen
  \bibfield  {author} {\bibinfo {author} {\bibfnamefont {R.}~\bibnamefont
  {Or\'us}}, \bibinfo {author} {\bibfnamefont {T.-C.}\ \bibnamefont {Wei}},
  \bibinfo {author} {\bibfnamefont {O.}~\bibnamefont {Buerschaper}}, \ and\
  \bibinfo {author} {\bibfnamefont {A.}~\bibnamefont {Garc\'{\i}a-Saez}},\
  }\href {\doibase 10.1103/PhysRevLett.113.257202} {\bibfield  {journal}
  {\bibinfo  {journal} {Phys. Rev. Lett.}\ }\textbf {\bibinfo {volume} {113}},\
  \bibinfo {pages} {257202} (\bibinfo {year} {2014})}\BibitemShut {NoStop}%
\bibitem [{\citenamefont {Verstraete}\ \emph {et~al.}(2004)\citenamefont
  {Verstraete}, \citenamefont {Garcia-Ripoll},\ and\ \citenamefont
  {Cirac}}]{Verstraete2004MPDO}%
  \BibitemOpen
  \bibfield  {author} {\bibinfo {author} {\bibfnamefont {F.}~\bibnamefont
  {Verstraete}}, \bibinfo {author} {\bibfnamefont {J.~J.}\ \bibnamefont
  {Garcia-Ripoll}}, \ and\ \bibinfo {author} {\bibfnamefont {J.~I.}\
  \bibnamefont {Cirac}},\ }\href {\doibase 10.1103/PhysRevLett.93.207204}
  {\bibfield  {journal} {\bibinfo  {journal} {Phys. Rev. Lett.}\ }\textbf
  {\bibinfo {volume} {93}},\ \bibinfo {pages} {207204} (\bibinfo {year}
  {2004})}\BibitemShut {NoStop}%
\bibitem [{\citenamefont {Werner}\ \emph {et~al.}(2016)\citenamefont {Werner},
  \citenamefont {Jaschke}, \citenamefont {Silvi}, \citenamefont {Kliesch},
  \citenamefont {Calarco}, \citenamefont {Eisert},\ and\ \citenamefont
  {Montangero}}]{Werner2016LPTN}%
  \BibitemOpen
  \bibfield  {author} {\bibinfo {author} {\bibfnamefont {A.~H.}\ \bibnamefont
  {Werner}}, \bibinfo {author} {\bibfnamefont {D.}~\bibnamefont {Jaschke}},
  \bibinfo {author} {\bibfnamefont {P.}~\bibnamefont {Silvi}}, \bibinfo
  {author} {\bibfnamefont {M.}~\bibnamefont {Kliesch}}, \bibinfo {author}
  {\bibfnamefont {T.}~\bibnamefont {Calarco}}, \bibinfo {author} {\bibfnamefont
  {J.}~\bibnamefont {Eisert}}, \ and\ \bibinfo {author} {\bibfnamefont
  {S.}~\bibnamefont {Montangero}},\ }\href {\doibase
  10.1103/PhysRevLett.116.237201} {\bibfield  {journal} {\bibinfo  {journal}
  {Physical Review Letters}\ }\textbf {\bibinfo {volume} {116}},\ \bibinfo
  {pages} {237201} (\bibinfo {year} {2016})}\BibitemShut {NoStop}%
\bibitem [{\citenamefont {Silvi}\ \emph {et~al.}(2014)\citenamefont {Silvi},
  \citenamefont {Rico}, \citenamefont {Calarco},\ and\ \citenamefont
  {Montangero}}]{Silvi2014LatticeGaugeTN}%
  \BibitemOpen
  \bibfield  {author} {\bibinfo {author} {\bibfnamefont {P.}~\bibnamefont
  {Silvi}}, \bibinfo {author} {\bibfnamefont {E.}~\bibnamefont {Rico}},
  \bibinfo {author} {\bibfnamefont {T.}~\bibnamefont {Calarco}}, \ and\
  \bibinfo {author} {\bibfnamefont {S.}~\bibnamefont {Montangero}},\ }\href
  {http://stacks.iop.org/1367-2630/16/i=10/a=103015} {\bibfield  {journal}
  {\bibinfo  {journal} {New Journal of Physics}\ }\textbf {\bibinfo {volume}
  {16}},\ \bibinfo {pages} {103015} (\bibinfo {year} {2014})}\BibitemShut
  {NoStop}%
\bibitem [{\citenamefont {Tagliacozzo}\ \emph {et~al.}(2014)\citenamefont
  {Tagliacozzo}, \citenamefont {Celi},\ and\ \citenamefont
  {Lewenstein}}]{Tagliacozzo2014LatticeGaugeTN}%
  \BibitemOpen
  \bibfield  {author} {\bibinfo {author} {\bibfnamefont {L.}~\bibnamefont
  {Tagliacozzo}}, \bibinfo {author} {\bibfnamefont {A.}~\bibnamefont {Celi}}, \
  and\ \bibinfo {author} {\bibfnamefont {M.}~\bibnamefont {Lewenstein}},\
  }\href {\doibase 10.1103/PhysRevX.4.041024} {\bibfield  {journal} {\bibinfo
  {journal} {Phys. Rev. X}\ }\textbf {\bibinfo {volume} {4}},\ \bibinfo {pages}
  {041024} (\bibinfo {year} {2014})}\BibitemShut {NoStop}%
\bibitem [{\citenamefont {Pichler}\ \emph {et~al.}(2016)\citenamefont
  {Pichler}, \citenamefont {Dalmonte}, \citenamefont {Rico}, \citenamefont
  {Zoller},\ and\ \citenamefont {Montangero}}]{Pichler2016U1LatticeGauge}%
  \BibitemOpen
  \bibfield  {author} {\bibinfo {author} {\bibfnamefont {T.}~\bibnamefont
  {Pichler}}, \bibinfo {author} {\bibfnamefont {M.}~\bibnamefont {Dalmonte}},
  \bibinfo {author} {\bibfnamefont {E.}~\bibnamefont {Rico}}, \bibinfo {author}
  {\bibfnamefont {P.}~\bibnamefont {Zoller}}, \ and\ \bibinfo {author}
  {\bibfnamefont {S.}~\bibnamefont {Montangero}},\ }\href {\doibase
  10.1103/PhysRevX.6.011023} {\bibfield  {journal} {\bibinfo  {journal} {Phys.
  Rev. X}\ }\textbf {\bibinfo {volume} {6}},\ \bibinfo {pages} {011023}
  (\bibinfo {year} {2016})}\BibitemShut {NoStop}%
\bibitem [{\citenamefont {Singh}\ \emph {et~al.}(2010)\citenamefont {Singh},
  \citenamefont {Pfeifer},\ and\ \citenamefont {Vidal}}]{Singh2010SymTN}%
  \BibitemOpen
  \bibfield  {author} {\bibinfo {author} {\bibfnamefont {S.}~\bibnamefont
  {Singh}}, \bibinfo {author} {\bibfnamefont {R.~N.~C.}\ \bibnamefont
  {Pfeifer}}, \ and\ \bibinfo {author} {\bibfnamefont {G.}~\bibnamefont
  {Vidal}},\ }\href {\doibase 10.1103/physreva.82.050301} {\bibfield  {journal}
  {\bibinfo  {journal} {Physical Review A}\ }\textbf {\bibinfo {volume} {82}},\
  \bibinfo {pages} {050301} (\bibinfo {year} {2010})}\BibitemShut {NoStop}%
\bibitem [{\citenamefont {Singh}\ \emph {et~al.}(2011)\citenamefont {Singh},
  \citenamefont {Pfeifer},\ and\ \citenamefont {Vidal}}]{Singh2011SymU1}%
  \BibitemOpen
  \bibfield  {author} {\bibinfo {author} {\bibfnamefont {S.}~\bibnamefont
  {Singh}}, \bibinfo {author} {\bibfnamefont {R.~N.~C.}\ \bibnamefont
  {Pfeifer}}, \ and\ \bibinfo {author} {\bibfnamefont {G.}~\bibnamefont
  {Vidal}},\ }\href {\doibase 10.1103/physrevb.83.115125} {\bibfield  {journal}
  {\bibinfo  {journal} {Physical Review B}\ }\textbf {\bibinfo {volume} {83}},\
  \bibinfo {pages} {115125} (\bibinfo {year} {2011})}\BibitemShut {NoStop}%
\bibitem [{\citenamefont {Singh}\ and\ \citenamefont
  {Vidal}(2012)}]{Singh2012SymSU2}%
  \BibitemOpen
  \bibfield  {author} {\bibinfo {author} {\bibfnamefont {S.}~\bibnamefont
  {Singh}}\ and\ \bibinfo {author} {\bibfnamefont {G.}~\bibnamefont {Vidal}},\
  }\href {\doibase 10.1103/physrevb.86.195114} {\bibfield  {journal} {\bibinfo
  {journal} {Physical Review B}\ }\textbf {\bibinfo {volume} {86}},\ \bibinfo
  {pages} {195114} (\bibinfo {year} {2012})}\BibitemShut {NoStop}%
\bibitem [{\citenamefont {Weichselbaum}(2012)}]{Weichselbaum2012SymNonabelian}%
  \BibitemOpen
  \bibfield  {author} {\bibinfo {author} {\bibfnamefont {A.}~\bibnamefont
  {Weichselbaum}},\ }\href {\doibase 10.1016/j.aop.2012.07.009} {\bibfield
  {journal} {\bibinfo  {journal} {Annals of Physics}\ }\textbf {\bibinfo
  {volume} {327}},\ \bibinfo {pages} {2972} (\bibinfo {year}
  {2012})}\BibitemShut {NoStop}%
\bibitem [{\citenamefont {Vidal}(2003)}]{Vidal2003TEBD}%
  \BibitemOpen
  \bibfield  {author} {\bibinfo {author} {\bibfnamefont {G.}~\bibnamefont
  {Vidal}},\ }\href {\doibase 10.1103/physrevlett.91.147902} {\bibfield
  {journal} {\bibinfo  {journal} {Physical Review Letters}\ }\textbf {\bibinfo
  {volume} {91}},\ \bibinfo {pages} {147902} (\bibinfo {year}
  {2003})}\BibitemShut {NoStop}%
\bibitem [{\citenamefont {Vidal}(2004)}]{Vidal2004TEBD}%
  \BibitemOpen
  \bibfield  {author} {\bibinfo {author} {\bibfnamefont {G.}~\bibnamefont
  {Vidal}},\ }\href {\doibase 10.1103/physrevlett.93.040502} {\bibfield
  {journal} {\bibinfo  {journal} {Physical Review Letters}\ }\textbf {\bibinfo
  {volume} {93}},\ \bibinfo {pages} {040502} (\bibinfo {year}
  {2004})}\BibitemShut {NoStop}%
\bibitem [{\citenamefont {Levin}\ and\ \citenamefont
  {Nave}(2007)}]{Levin2007TRG}%
  \BibitemOpen
  \bibfield  {author} {\bibinfo {author} {\bibfnamefont {M.}~\bibnamefont
  {Levin}}\ and\ \bibinfo {author} {\bibfnamefont {C.~P.}\ \bibnamefont
  {Nave}},\ }\href {\doibase 10.1103/physrevlett.99.120601} {\bibfield
  {journal} {\bibinfo  {journal} {Physical Review Letters}\ }\textbf {\bibinfo
  {volume} {99}},\ \bibinfo {pages} {120601} (\bibinfo {year}
  {2007})}\BibitemShut {NoStop}%
\bibitem [{\citenamefont {Nishino}\ and\ \citenamefont
  {Okunishi}(1996)}]{Nishino1996CTMRG}%
  \BibitemOpen
  \bibfield  {author} {\bibinfo {author} {\bibfnamefont {T.}~\bibnamefont
  {Nishino}}\ and\ \bibinfo {author} {\bibfnamefont {K.}~\bibnamefont
  {Okunishi}},\ }\href {\doibase 10.1143/jpsj.65.891} {\bibfield  {journal}
  {\bibinfo  {journal} {Journal of the Physical Society of Japan}\ }\textbf
  {\bibinfo {volume} {65}},\ \bibinfo {pages} {891} (\bibinfo {year}
  {1996})}\BibitemShut {NoStop}%
\bibitem [{\citenamefont {Verstraete}\ and\ \citenamefont
  {Cirac}(2004)}]{Verstraete2004PEPS}%
  \BibitemOpen
  \bibfield  {author} {\bibinfo {author} {\bibfnamefont {F.}~\bibnamefont
  {Verstraete}}\ and\ \bibinfo {author} {\bibfnamefont {J.~I.}\ \bibnamefont
  {Cirac}},\ }\href@noop {} {\  (\bibinfo {year} {2004})},\ \Eprint
  {http://arxiv.org/abs/arXiv:cond-mat/0407066} {arXiv:cond-mat/0407066}
  \BibitemShut {NoStop}%
\bibitem [{\citenamefont {Murg}\ \emph {et~al.}(2007)\citenamefont {Murg},
  \citenamefont {Verstraete},\ and\ \citenamefont {Cirac}}]{Murg2007PEPS2}%
  \BibitemOpen
  \bibfield  {author} {\bibinfo {author} {\bibfnamefont {V.}~\bibnamefont
  {Murg}}, \bibinfo {author} {\bibfnamefont {F.}~\bibnamefont {Verstraete}}, \
  and\ \bibinfo {author} {\bibfnamefont {J.~I.}\ \bibnamefont {Cirac}},\ }\href
  {\doibase 10.1103/physreva.75.033605} {\bibfield  {journal} {\bibinfo
  {journal} {Physical Review A}\ }\textbf {\bibinfo {volume} {75}},\ \bibinfo
  {pages} {033605} (\bibinfo {year} {2007})}\BibitemShut {NoStop}%
\bibitem [{\citenamefont {Golub}\ and\ \citenamefont
  {Van~Loan}(2012)}]{Golub2012MatrixComputations}%
  \BibitemOpen
  \bibfield  {author} {\bibinfo {author} {\bibfnamefont {G.~H.}\ \bibnamefont
  {Golub}}\ and\ \bibinfo {author} {\bibfnamefont {C.~F.}\ \bibnamefont
  {Van~Loan}},\ }\href@noop {} {\emph {\bibinfo {title} {Matrix
  computations}}},\ Vol.~\bibinfo {volume} {3}\ (\bibinfo  {publisher} {Johns
  Hopkins University Press},\ \bibinfo {address} {Baltimore, MA, USA},\
  \bibinfo {year} {2012})\BibitemShut {NoStop}%
\bibitem [{\citenamefont {Demmel}(1997)}]{Demmel1997NumericalLA}%
  \BibitemOpen
  \bibfield  {author} {\bibinfo {author} {\bibfnamefont {J.~W.}\ \bibnamefont
  {Demmel}},\ }\href {\doibase 10.1137/1.9781611971446} {\emph {\bibinfo
  {title} {Applied Numerical Linear Algebra}}}\ (\bibinfo  {publisher} {Society
  for Industrial and Applied Mathematics},\ \bibinfo {year} {1997})\BibitemShut
  {NoStop}%
\bibitem [{\citenamefont {Erichson}\ \emph {et~al.}(2017)\citenamefont
  {Erichson}, \citenamefont {Manohar}, \citenamefont {Brunton},\ and\
  \citenamefont {Kutz}}]{Erichson2017RandomCPTensorDecomp}%
  \BibitemOpen
  \bibfield  {author} {\bibinfo {author} {\bibfnamefont {N.~B.}\ \bibnamefont
  {Erichson}}, \bibinfo {author} {\bibfnamefont {K.}~\bibnamefont {Manohar}},
  \bibinfo {author} {\bibfnamefont {S.~L.}\ \bibnamefont {Brunton}}, \ and\
  \bibinfo {author} {\bibfnamefont {J.~N.}\ \bibnamefont {Kutz}},\ }\href@noop
  {} {\  (\bibinfo {year} {2017})},\ \Eprint
  {http://arxiv.org/abs/arXiv:1703.09074} {arXiv:1703.09074} \BibitemShut
  {NoStop}%
\bibitem [{\citenamefont {Hastie}\ \emph {et~al.}(2009)\citenamefont {Hastie},
  \citenamefont {Tibshirani},\ and\ \citenamefont
  {Friedman}}]{Hastie2009StatisticalLearning}%
  \BibitemOpen
  \bibfield  {author} {\bibinfo {author} {\bibfnamefont {T.}~\bibnamefont
  {Hastie}}, \bibinfo {author} {\bibfnamefont {R.}~\bibnamefont {Tibshirani}},
  \ and\ \bibinfo {author} {\bibfnamefont {J.}~\bibnamefont {Friedman}},\
  }\href {\doibase 10.1007/978-0-387-84858-7} {\emph {\bibinfo {title} {The
  Elements of Statistical Learning: Data Mining, Inference, and Prediction}}},\
  \bibinfo {edition} {2nd}\ ed.\ (\bibinfo  {publisher} {Springer New York},\
  \bibinfo {year} {2009})\BibitemShut {NoStop}%
\bibitem [{\citenamefont {Cichocki}\ \emph {et~al.}(2015)\citenamefont
  {Cichocki}, \citenamefont {Mandic}, \citenamefont {Phan}, \citenamefont
  {Caiafa}, \citenamefont {Zhou}, \citenamefont {Zhao},\ and\ \citenamefont
  {Lathauwer}}]{Cichocki2015SignalProcessingTensors}%
  \BibitemOpen
  \bibfield  {author} {\bibinfo {author} {\bibfnamefont {A.}~\bibnamefont
  {Cichocki}}, \bibinfo {author} {\bibfnamefont {D.~P.}\ \bibnamefont
  {Mandic}}, \bibinfo {author} {\bibfnamefont {A.~H.}\ \bibnamefont {Phan}},
  \bibinfo {author} {\bibfnamefont {C.~F.}\ \bibnamefont {Caiafa}}, \bibinfo
  {author} {\bibfnamefont {G.}~\bibnamefont {Zhou}}, \bibinfo {author}
  {\bibfnamefont {Q.}~\bibnamefont {Zhao}}, \ and\ \bibinfo {author}
  {\bibfnamefont {L.~D.}\ \bibnamefont {Lathauwer}},\ }\href {\doibase
  10.1109/MSP.2013.2297439} {\bibfield  {journal} {\bibinfo  {journal} {IEEE
  Signal Processing Magazine}\ }\textbf {\bibinfo {volume} {32}},\ \bibinfo
  {pages} {145} (\bibinfo {year} {2015})}\BibitemShut {NoStop}%
\bibitem [{\citenamefont
  {Vijayakumari}(2013)}]{Vijayakumari2013FaceRecognitionSurvey}%
  \BibitemOpen
  \bibfield  {author} {\bibinfo {author} {\bibfnamefont {V.}~\bibnamefont
  {Vijayakumari}},\ }\href@noop {} {\bibfield  {journal} {\bibinfo  {journal}
  {World Journal of Computer Application and Technology}\ }\textbf {\bibinfo
  {volume} {1}},\ \bibinfo {pages} {41} (\bibinfo {year} {2013})}\BibitemShut
  {NoStop}%
\bibitem [{\citenamefont {Pai}\ \emph {et~al.}(2015)\citenamefont {Pai},
  \citenamefont {Fernandes}, \citenamefont {Nayak}, \citenamefont {Nagesha},
  \citenamefont {Accamma}, \citenamefont {Sushmitha},\ and\ \citenamefont
  {Kumari}}]{Pai2015FaceRecognitionIllumination}%
  \BibitemOpen
  \bibfield  {author} {\bibinfo {author} {\bibfnamefont {A.~G.}\ \bibnamefont
  {Pai}}, \bibinfo {author} {\bibfnamefont {S.~L.}\ \bibnamefont {Fernandes}},
  \bibinfo {author} {\bibfnamefont {K.}~\bibnamefont {Nayak}}, \bibinfo
  {author} {\bibnamefont {Nagesha}}, \bibinfo {author} {\bibfnamefont {K.~K.}\
  \bibnamefont {Accamma}}, \bibinfo {author} {\bibfnamefont {K.}~\bibnamefont
  {Sushmitha}}, \ and\ \bibinfo {author} {\bibfnamefont {K.}~\bibnamefont
  {Kumari}},\ }in\ \href {\doibase 10.1109/ECS.2015.7124974} {\emph {\bibinfo
  {booktitle} {2nd International Conference on Electronics and Communication
  Systems (ICECS)}}}\ (\bibinfo {organization} {IEEE},\ \bibinfo {year}
  {2015})\ pp.\ \bibinfo {pages} {577--582}\BibitemShut {NoStop}%
\bibitem [{\citenamefont {Wall}\ \emph {et~al.}(2003)\citenamefont {Wall},
  \citenamefont {Rechtsteiner},\ and\ \citenamefont
  {Rocha}}]{Wall2003MicroarrayAnalysisSVD}%
  \BibitemOpen
  \bibfield  {author} {\bibinfo {author} {\bibfnamefont {M.~E.}\ \bibnamefont
  {Wall}}, \bibinfo {author} {\bibfnamefont {A.}~\bibnamefont {Rechtsteiner}},
  \ and\ \bibinfo {author} {\bibfnamefont {L.~M.}\ \bibnamefont {Rocha}},\ }in\
  \href@noop {} {\emph {\bibinfo {booktitle} {A Practical Approach to
  Microarray Data Analysis}}},\ \bibinfo {editor} {edited by\ \bibinfo {editor}
  {\bibfnamefont {G.~M.}\ \bibnamefont {Berrar~D.P.}, \bibfnamefont
  {Dubitzky~W.}}}\ (\bibinfo  {publisher} {Springer},\ \bibinfo {year} {2003})\
  pp.\ \bibinfo {pages} {91--109}\BibitemShut {NoStop}%
\bibitem [{\citenamefont {Jermyn}(2017)}]{Jermyn2017automatic}%
  \BibitemOpen
  \bibfield  {author} {\bibinfo {author} {\bibfnamefont {A.~S.}\ \bibnamefont
  {Jermyn}},\ }\href@noop {} {\  (\bibinfo {year} {2017})},\ \Eprint
  {http://arxiv.org/abs/arXiv:1709.03080} {arXiv:1709.03080} \BibitemShut
  {NoStop}%
\bibitem [{\citenamefont {Halko}\ \emph {et~al.}(2011)\citenamefont {Halko},
  \citenamefont {Martinsson},\ and\ \citenamefont
  {Tropp}}]{Halko2011LowRankProbabilistic}%
  \BibitemOpen
  \bibfield  {author} {\bibinfo {author} {\bibfnamefont {N.}~\bibnamefont
  {Halko}}, \bibinfo {author} {\bibfnamefont {P.-G.}\ \bibnamefont
  {Martinsson}}, \ and\ \bibinfo {author} {\bibfnamefont {J.~A.}\ \bibnamefont
  {Tropp}},\ }\href {\doibase 10.1137/090771806} {\bibfield  {journal}
  {\bibinfo  {journal} {{SIAM} Review}\ }\textbf {\bibinfo {volume} {53}},\
  \bibinfo {pages} {217} (\bibinfo {year} {2011})}\BibitemShut {NoStop}%
\bibitem [{\citenamefont {Tamascelli}\ \emph {et~al.}(2015)\citenamefont
  {Tamascelli}, \citenamefont {Rosenbach},\ and\ \citenamefont
  {Plenio}}]{Tamascelli2015TEBDRSVD}%
  \BibitemOpen
  \bibfield  {author} {\bibinfo {author} {\bibfnamefont {D.}~\bibnamefont
  {Tamascelli}}, \bibinfo {author} {\bibfnamefont {R.}~\bibnamefont
  {Rosenbach}}, \ and\ \bibinfo {author} {\bibfnamefont {M.~B.}\ \bibnamefont
  {Plenio}},\ }\href {\doibase 10.1103/physreve.91.063306} {\bibfield
  {journal} {\bibinfo  {journal} {Physical Review E}\ }\textbf {\bibinfo
  {volume} {91}},\ \bibinfo {pages} {063306} (\bibinfo {year}
  {2015})}\BibitemShut {NoStop}%
\bibitem [{\citenamefont {Anderson}\ \emph {et~al.}(1999)\citenamefont
  {Anderson}, \citenamefont {Bai}, \citenamefont {Bischof}, \citenamefont
  {Blackford}, \citenamefont {Demmel}, \citenamefont {Dongarra}, \citenamefont
  {Du~Croz}, \citenamefont {Greenbaum}, \citenamefont {Hammarling},
  \citenamefont {McKenney},\ and\ \citenamefont
  {Sorensen}}]{Anderson1999LAPACK_userguide}%
  \BibitemOpen
  \bibfield  {author} {\bibinfo {author} {\bibfnamefont {E.}~\bibnamefont
  {Anderson}}, \bibinfo {author} {\bibfnamefont {Z.}~\bibnamefont {Bai}},
  \bibinfo {author} {\bibfnamefont {C.}~\bibnamefont {Bischof}}, \bibinfo
  {author} {\bibfnamefont {L.}~\bibnamefont {Blackford}}, \bibinfo {author}
  {\bibfnamefont {J.}~\bibnamefont {Demmel}}, \bibinfo {author} {\bibfnamefont
  {J.}~\bibnamefont {Dongarra}}, \bibinfo {author} {\bibfnamefont
  {J.}~\bibnamefont {Du~Croz}}, \bibinfo {author} {\bibfnamefont
  {A.}~\bibnamefont {Greenbaum}}, \bibinfo {author} {\bibfnamefont
  {S.}~\bibnamefont {Hammarling}}, \bibinfo {author} {\bibfnamefont
  {A.}~\bibnamefont {McKenney}}, \ and\ \bibinfo {author} {\bibfnamefont
  {D.}~\bibnamefont {Sorensen}},\ }\href {\doibase 10.1137/1.9780898719604}
  {\emph {\bibinfo {title} {{LAPACK} Users' Guide}}},\ \bibinfo {edition}
  {3rd}\ ed.\ (\bibinfo  {publisher} {Society for Industrial and Applied
  Mathematics},\ \bibinfo {year} {1999})\BibitemShut {NoStop}%
\bibitem [{\citenamefont {Shi}\ \emph {et~al.}(2006)\citenamefont {Shi},
  \citenamefont {Duan},\ and\ \citenamefont {Vidal}}]{Shi2006TTN}%
  \BibitemOpen
  \bibfield  {author} {\bibinfo {author} {\bibfnamefont {Y.-Y.}\ \bibnamefont
  {Shi}}, \bibinfo {author} {\bibfnamefont {L.-M.}\ \bibnamefont {Duan}}, \
  and\ \bibinfo {author} {\bibfnamefont {G.}~\bibnamefont {Vidal}},\ }\href
  {\doibase 10.1103/physreva.74.022320} {\bibfield  {journal} {\bibinfo
  {journal} {Physical Review A}\ }\textbf {\bibinfo {volume} {74}},\ \bibinfo
  {pages} {022320} (\bibinfo {year} {2006})}\BibitemShut {NoStop}%
\bibitem [{\citenamefont {Hackbusch}\ and\ \citenamefont
  {K{\"u}hn}(2009)}]{Hackbusch2009TTNScheme}%
  \BibitemOpen
  \bibfield  {author} {\bibinfo {author} {\bibfnamefont {W.}~\bibnamefont
  {Hackbusch}}\ and\ \bibinfo {author} {\bibfnamefont {S.}~\bibnamefont
  {K{\"u}hn}},\ }\href {\doibase 10.1007/s00041-009-9094-9} {\bibfield
  {journal} {\bibinfo  {journal} {Journal of Fourier Analysis and
  Applications}\ }\textbf {\bibinfo {volume} {15}},\ \bibinfo {pages} {706}
  (\bibinfo {year} {2009})}\BibitemShut {NoStop}%
\bibitem [{\citenamefont {Murg}\ \emph {et~al.}(2010)\citenamefont {Murg},
  \citenamefont {Verstraete}, \citenamefont {Legeza},\ and\ \citenamefont
  {Noack}}]{Murg2010TTN}%
  \BibitemOpen
  \bibfield  {author} {\bibinfo {author} {\bibfnamefont {V.}~\bibnamefont
  {Murg}}, \bibinfo {author} {\bibfnamefont {F.}~\bibnamefont {Verstraete}},
  \bibinfo {author} {\bibfnamefont {O.}~\bibnamefont {Legeza}}, \ and\ \bibinfo
  {author} {\bibfnamefont {R.~M.}\ \bibnamefont {Noack}},\ }\href {\doibase
  10.1103/physrevb.82.205105} {\bibfield  {journal} {\bibinfo  {journal}
  {Physical Review B}\ }\textbf {\bibinfo {volume} {82}},\ \bibinfo {pages}
  {205105} (\bibinfo {year} {2010})}\BibitemShut {NoStop}%
\bibitem [{\citenamefont {Nakatani}\ and\ \citenamefont
  {Chan}(2013)}]{Nakatani2013TTN}%
  \BibitemOpen
  \bibfield  {author} {\bibinfo {author} {\bibfnamefont {N.}~\bibnamefont
  {Nakatani}}\ and\ \bibinfo {author} {\bibfnamefont {G.~K.-L.}\ \bibnamefont
  {Chan}},\ }\href {\doibase 10.1063/1.4798639} {\bibfield  {journal} {\bibinfo
   {journal} {The Journal of Chemical Physics}\ }\textbf {\bibinfo {volume}
  {138}},\ \bibinfo {pages} {134113} (\bibinfo {year} {2013})}\BibitemShut
  {NoStop}%
\bibitem [{\citenamefont {Gerster}\ \emph {et~al.}(2014)\citenamefont
  {Gerster}, \citenamefont {Silvi}, \citenamefont {Rizzi}, \citenamefont
  {Fazio}, \citenamefont {Calarco},\ and\ \citenamefont
  {Montangero}}]{Gerster2014TTN}%
  \BibitemOpen
  \bibfield  {author} {\bibinfo {author} {\bibfnamefont {M.}~\bibnamefont
  {Gerster}}, \bibinfo {author} {\bibfnamefont {P.}~\bibnamefont {Silvi}},
  \bibinfo {author} {\bibfnamefont {M.}~\bibnamefont {Rizzi}}, \bibinfo
  {author} {\bibfnamefont {R.}~\bibnamefont {Fazio}}, \bibinfo {author}
  {\bibfnamefont {T.}~\bibnamefont {Calarco}}, \ and\ \bibinfo {author}
  {\bibfnamefont {S.}~\bibnamefont {Montangero}},\ }\href {\doibase
  10.1103/physrevb.90.125154} {\bibfield  {journal} {\bibinfo  {journal}
  {Physical Review B}\ }\textbf {\bibinfo {volume} {90}},\ \bibinfo {pages}
  {125154} (\bibinfo {year} {2014})}\BibitemShut {NoStop}%
\bibitem [{\citenamefont {Gerster}\ \emph {et~al.}(2016)\citenamefont
  {Gerster}, \citenamefont {Rizzi}, \citenamefont {Tschirsich}, \citenamefont
  {Silvi}, \citenamefont {Fazio},\ and\ \citenamefont
  {Montangero}}]{Gerster2016BoseHubbard}%
  \BibitemOpen
  \bibfield  {author} {\bibinfo {author} {\bibfnamefont {M.}~\bibnamefont
  {Gerster}}, \bibinfo {author} {\bibfnamefont {M.}~\bibnamefont {Rizzi}},
  \bibinfo {author} {\bibfnamefont {F.}~\bibnamefont {Tschirsich}}, \bibinfo
  {author} {\bibfnamefont {P.}~\bibnamefont {Silvi}}, \bibinfo {author}
  {\bibfnamefont {R.}~\bibnamefont {Fazio}}, \ and\ \bibinfo {author}
  {\bibfnamefont {S.}~\bibnamefont {Montangero}},\ }\href
  {http://stacks.iop.org/1367-2630/18/i=1/a=015015} {\bibfield  {journal}
  {\bibinfo  {journal} {New Journal of Physics}\ }\textbf {\bibinfo {volume}
  {18}},\ \bibinfo {pages} {015015} (\bibinfo {year} {2016})}\BibitemShut
  {NoStop}%
\bibitem [{\citenamefont {Eckart}\ and\ \citenamefont
  {Young}(1936)}]{Eckart1936LowRank}%
  \BibitemOpen
  \bibfield  {author} {\bibinfo {author} {\bibfnamefont {C.}~\bibnamefont
  {Eckart}}\ and\ \bibinfo {author} {\bibfnamefont {G.}~\bibnamefont {Young}},\
  }\href {\doibase 10.1007/bf02288367} {\bibfield  {journal} {\bibinfo
  {journal} {Psychometrika}\ }\textbf {\bibinfo {volume} {1}},\ \bibinfo
  {pages} {211} (\bibinfo {year} {1936})}\BibitemShut {NoStop}%
\bibitem [{\citenamefont {Mirsky}(1960)}]{Mirsky1960Symmetric}%
  \BibitemOpen
  \bibfield  {author} {\bibinfo {author} {\bibfnamefont {L.}~\bibnamefont
  {Mirsky}},\ }\href@noop {} {\bibfield  {journal} {\bibinfo  {journal} {The
  quarterly journal of mathematics}\ }\textbf {\bibinfo {volume} {11}},\
  \bibinfo {pages} {50} (\bibinfo {year} {1960})}\BibitemShut {NoStop}%
\bibitem [{\citenamefont {Martinsson}\ \emph {et~al.}(2011)\citenamefont
  {Martinsson}, \citenamefont {Rokhlin},\ and\ \citenamefont
  {Tygert}}]{Martinsson2011RandLowRankNoPI}%
  \BibitemOpen
  \bibfield  {author} {\bibinfo {author} {\bibfnamefont {P.-G.}\ \bibnamefont
  {Martinsson}}, \bibinfo {author} {\bibfnamefont {V.}~\bibnamefont {Rokhlin}},
  \ and\ \bibinfo {author} {\bibfnamefont {M.}~\bibnamefont {Tygert}},\ }\href
  {\doibase 10.1016/j.acha.2010.02.003} {\bibfield  {journal} {\bibinfo
  {journal} {Applied and Computational Harmonic Analysis}\ }\textbf {\bibinfo
  {volume} {30}},\ \bibinfo {pages} {47} (\bibinfo {year} {2011})}\BibitemShut
  {NoStop}%
\bibitem [{\citenamefont {Rokhlin}\ \emph {et~al.}(2010)\citenamefont
  {Rokhlin}, \citenamefont {Szlam},\ and\ \citenamefont
  {Tygert}}]{Rokhlin2010LowRankPIScheme}%
  \BibitemOpen
  \bibfield  {author} {\bibinfo {author} {\bibfnamefont {V.}~\bibnamefont
  {Rokhlin}}, \bibinfo {author} {\bibfnamefont {A.}~\bibnamefont {Szlam}}, \
  and\ \bibinfo {author} {\bibfnamefont {M.}~\bibnamefont {Tygert}},\ }\href
  {\doibase 10.1137/080736417} {\bibfield  {journal} {\bibinfo  {journal}
  {{SIAM} Journal on Matrix Analysis and Applications}\ }\textbf {\bibinfo
  {volume} {31}},\ \bibinfo {pages} {1100} (\bibinfo {year}
  {2010})}\BibitemShut {NoStop}%
\bibitem [{\citenamefont {Gu}(2015)}]{Gu2015LowRankPIError}%
  \BibitemOpen
  \bibfield  {author} {\bibinfo {author} {\bibfnamefont {M.}~\bibnamefont
  {Gu}},\ }\href {\doibase 10.1137/130938700} {\bibfield  {journal} {\bibinfo
  {journal} {{SIAM} Journal on Scientific Computing}\ }\textbf {\bibinfo
  {volume} {37}},\ \bibinfo {pages} {A1139} (\bibinfo {year}
  {2015})}\BibitemShut {NoStop}%
\bibitem [{\citenamefont {Szlam}\ \emph {et~al.}(2014)\citenamefont {Szlam},
  \citenamefont {Kluger},\ and\ \citenamefont
  {Tygert}}]{Szlam2014RSVDinMATLAB}%
  \BibitemOpen
  \bibfield  {author} {\bibinfo {author} {\bibfnamefont {A.}~\bibnamefont
  {Szlam}}, \bibinfo {author} {\bibfnamefont {Y.}~\bibnamefont {Kluger}}, \
  and\ \bibinfo {author} {\bibfnamefont {M.}~\bibnamefont {Tygert}},\
  }\href@noop {} {\  (\bibinfo {year} {2014})},\ \Eprint
  {http://arxiv.org/abs/arXiv:1412.3510} {arXiv:1412.3510} \BibitemShut
  {NoStop}%
\bibitem [{\citenamefont {Erichson}\ \emph {et~al.}(2016)\citenamefont
  {Erichson}, \citenamefont {Voronin}, \citenamefont {Brunton},\ and\
  \citenamefont {Kutz}}]{Erichson2016LowRankProbabilisticInR}%
  \BibitemOpen
  \bibfield  {author} {\bibinfo {author} {\bibfnamefont {N.~B.}\ \bibnamefont
  {Erichson}}, \bibinfo {author} {\bibfnamefont {S.}~\bibnamefont {Voronin}},
  \bibinfo {author} {\bibfnamefont {S.~L.}\ \bibnamefont {Brunton}}, \ and\
  \bibinfo {author} {\bibfnamefont {J.~N.}\ \bibnamefont {Kutz}},\ }\href@noop
  {} {\  (\bibinfo {year} {2016})},\ \Eprint
  {http://arxiv.org/abs/arXiv:1608.02148} {arXiv:1608.02148} \BibitemShut
  {NoStop}%
\bibitem [{\citenamefont {Voronin}\ and\ \citenamefont
  {Martinsson}(2015)}]{Voronin2015RSVDPACK}%
  \BibitemOpen
  \bibfield  {author} {\bibinfo {author} {\bibfnamefont {S.}~\bibnamefont
  {Voronin}}\ and\ \bibinfo {author} {\bibfnamefont {P.-G.}\ \bibnamefont
  {Martinsson}},\ }\href@noop {} {\  (\bibinfo {year} {2015})},\ \Eprint
  {http://arxiv.org/abs/arXiv:1502.05366} {arXiv:1502.05366} \BibitemShut
  {NoStop}%
\bibitem [{\citenamefont {Silvi}\ \emph {et~al.}(tion)\citenamefont {Silvi},
  \citenamefont {Tschirsich}, \citenamefont {Gerster}, \citenamefont
  {J\"unemann}, \citenamefont {Jaschke}, \citenamefont {Rizzi},\ and\
  \citenamefont {Montangero}}]{Silvi2018LoopfreeTNMethods}%
  \BibitemOpen
  \bibfield  {author} {\bibinfo {author} {\bibfnamefont {P.}~\bibnamefont
  {Silvi}}, \bibinfo {author} {\bibfnamefont {F.}~\bibnamefont {Tschirsich}},
  \bibinfo {author} {\bibfnamefont {M.}~\bibnamefont {Gerster}}, \bibinfo
  {author} {\bibfnamefont {J.}~\bibnamefont {J\"unemann}}, \bibinfo {author}
  {\bibfnamefont {D.}~\bibnamefont {Jaschke}}, \bibinfo {author} {\bibfnamefont
  {M.}~\bibnamefont {Rizzi}}, \ and\ \bibinfo {author} {\bibfnamefont
  {S.}~\bibnamefont {Montangero}},\ }\href@noop {} {\  (\bibinfo {year} {in
  preparation})}\BibitemShut {NoStop}%
\bibitem [{\citenamefont {Penson}\ and\ \citenamefont
  {Kolb}(1984)}]{Penson1984TransvIsingHighSpinCritical}%
  \BibitemOpen
  \bibfield  {author} {\bibinfo {author} {\bibfnamefont {K.~A.}\ \bibnamefont
  {Penson}}\ and\ \bibinfo {author} {\bibfnamefont {M.}~\bibnamefont {Kolb}},\
  }\href {\doibase 10.1103/PhysRevB.30.1470} {\bibfield  {journal} {\bibinfo
  {journal} {Phys. Rev. B}\ }\textbf {\bibinfo {volume} {30}},\ \bibinfo
  {pages} {1470} (\bibinfo {year} {1984})}\BibitemShut {NoStop}%
\bibitem [{\citenamefont {Fisher}\ and\ \citenamefont
  {Barber}(1972)}]{Fisher1972FSScaling}%
  \BibitemOpen
  \bibfield  {author} {\bibinfo {author} {\bibfnamefont {M.~E.}\ \bibnamefont
  {Fisher}}\ and\ \bibinfo {author} {\bibfnamefont {M.~N.}\ \bibnamefont
  {Barber}},\ }\href@noop {} {\bibfield  {journal} {\bibinfo  {journal}
  {Physical Review Letters}\ }\textbf {\bibinfo {volume} {28}},\ \bibinfo
  {pages} {1516} (\bibinfo {year} {1972})}\BibitemShut {NoStop}%
\bibitem [{\citenamefont {du~Croo~de Jongh}\ and\ \citenamefont {van
  Leeuwen}(1998)}]{Jongh1998Ising2DDMRG}%
  \BibitemOpen
  \bibfield  {author} {\bibinfo {author} {\bibfnamefont {M.~S.~L.}\
  \bibnamefont {du~Croo~de Jongh}}\ and\ \bibinfo {author} {\bibfnamefont
  {J.~M.~J.}\ \bibnamefont {van Leeuwen}},\ }\href {\doibase
  10.1103/physrevb.57.8494} {\bibfield  {journal} {\bibinfo  {journal}
  {Physical Review B}\ }\textbf {\bibinfo {volume} {57}},\ \bibinfo {pages}
  {8494} (\bibinfo {year} {1998})}\BibitemShut {NoStop}%
\bibitem [{\citenamefont {Bl{\"o}te}\ and\ \citenamefont
  {Deng}(2002)}]{Bloete2002IsingMC}%
  \BibitemOpen
  \bibfield  {author} {\bibinfo {author} {\bibfnamefont {H.~W.}\ \bibnamefont
  {Bl{\"o}te}}\ and\ \bibinfo {author} {\bibfnamefont {Y.}~\bibnamefont
  {Deng}},\ }\href {\doibase 10.1103/physreve.66.066110} {\bibfield  {journal}
  {\bibinfo  {journal} {Physical Review E}\ }\textbf {\bibinfo {volume} {66}},\
  \bibinfo {pages} {066110} (\bibinfo {year} {2002})}\BibitemShut {NoStop}%
\bibitem [{\citenamefont {Rizzi}\ \emph {et~al.}(2010)\citenamefont {Rizzi},
  \citenamefont {Montangero}, \citenamefont {Silvi}, \citenamefont
  {Giovannetti},\ and\ \citenamefont {Fazio}}]{Rizzi2010MERACritical}%
  \BibitemOpen
  \bibfield  {author} {\bibinfo {author} {\bibfnamefont {M.}~\bibnamefont
  {Rizzi}}, \bibinfo {author} {\bibfnamefont {S.}~\bibnamefont {Montangero}},
  \bibinfo {author} {\bibfnamefont {P.}~\bibnamefont {Silvi}}, \bibinfo
  {author} {\bibfnamefont {V.}~\bibnamefont {Giovannetti}}, \ and\ \bibinfo
  {author} {\bibfnamefont {R.}~\bibnamefont {Fazio}},\ }\href
  {http://stacks.iop.org/1367-2630/12/i=7/a=075018} {\bibfield  {journal}
  {\bibinfo  {journal} {New Journal of Physics}\ }\textbf {\bibinfo {volume}
  {12}},\ \bibinfo {pages} {075018} (\bibinfo {year} {2010})}\BibitemShut
  {NoStop}%
\bibitem [{\citenamefont {Calabrese}\ and\ \citenamefont
  {Lefevre}(2008)}]{Calabrese2008CriticalSvals1D}%
  \BibitemOpen
  \bibfield  {author} {\bibinfo {author} {\bibfnamefont {P.}~\bibnamefont
  {Calabrese}}\ and\ \bibinfo {author} {\bibfnamefont {A.}~\bibnamefont
  {Lefevre}},\ }\href {\doibase 10.1103/physreva.78.032329} {\bibfield
  {journal} {\bibinfo  {journal} {Physical Review A}\ }\textbf {\bibinfo
  {volume} {78}},\ \bibinfo {pages} {032329} (\bibinfo {year}
  {2008})}\BibitemShut {NoStop}%
\bibitem [{\citenamefont {Cardy}(2012)}]{Cardy2012FSScaling}%
  \BibitemOpen
  \bibfield  {author} {\bibinfo {author} {\bibfnamefont {J.}~\bibnamefont
  {Cardy}},\ }\href@noop {} {\emph {\bibinfo {title} {Finite-size scaling}}},\
  Vol.~\bibinfo {volume} {2}\ (\bibinfo  {publisher} {Elsevier},\ \bibinfo
  {year} {2012})\BibitemShut {NoStop}%
\bibitem [{\citenamefont {Calabrese}\ and\ \citenamefont
  {Cardy}(2009)}]{Calabrese2009EntanglementEntropyReview}%
  \BibitemOpen
  \bibfield  {author} {\bibinfo {author} {\bibfnamefont {P.}~\bibnamefont
  {Calabrese}}\ and\ \bibinfo {author} {\bibfnamefont {J.}~\bibnamefont
  {Cardy}},\ }\href {http://stacks.iop.org/1751-8121/42/i=50/a=504005}
  {\bibfield  {journal} {\bibinfo  {journal} {Journal of Physics A:
  Mathematical and Theoretical}\ }\textbf {\bibinfo {volume} {42}},\ \bibinfo
  {pages} {504005} (\bibinfo {year} {2009})}\BibitemShut {NoStop}%
\bibitem [{\citenamefont {Musco}\ and\ \citenamefont
  {Musco}(2015)}]{Musco2015RandomizedKrylowSVD}%
  \BibitemOpen
  \bibfield  {author} {\bibinfo {author} {\bibfnamefont {C.}~\bibnamefont
  {Musco}}\ and\ \bibinfo {author} {\bibfnamefont {C.}~\bibnamefont {Musco}},\
  }in\ \href
  {http://papers.nips.cc/paper/5735-randomized-block-krylov-methods-for-stronger-and-faster-approximate-singular-value-decomposition.pdf}
  {\emph {\bibinfo {booktitle} {Advances in Neural Information Processing
  Systems 28}}},\ \bibinfo {editor} {edited by\ \bibinfo {editor}
  {\bibfnamefont {C.}~\bibnamefont {Cortes}}, \bibinfo {editor} {\bibfnamefont
  {N.~D.}\ \bibnamefont {Lawrence}}, \bibinfo {editor} {\bibfnamefont {D.~D.}\
  \bibnamefont {Lee}}, \bibinfo {editor} {\bibfnamefont {M.}~\bibnamefont
  {Sugiyama}}, \ and\ \bibinfo {editor} {\bibfnamefont {R.}~\bibnamefont
  {Garnett}}}\ (\bibinfo  {publisher} {Curran Associates, Inc.},\ \bibinfo
  {year} {2015})\ pp.\ \bibinfo {pages} {1396--1404}\BibitemShut {NoStop}%
\bibitem [{\citenamefont {Tropp}\ \emph {et~al.}(2016)\citenamefont {Tropp},
  \citenamefont {Yurtsever}, \citenamefont {Udell},\ and\ \citenamefont
  {Cevher}}]{Tropp2016SingleView}%
  \BibitemOpen
  \bibfield  {author} {\bibinfo {author} {\bibfnamefont {J.~A.}\ \bibnamefont
  {Tropp}}, \bibinfo {author} {\bibfnamefont {A.}~\bibnamefont {Yurtsever}},
  \bibinfo {author} {\bibfnamefont {M.}~\bibnamefont {Udell}}, \ and\ \bibinfo
  {author} {\bibfnamefont {V.}~\bibnamefont {Cevher}},\ }\href@noop {} {\
  (\bibinfo {year} {2016})},\ \Eprint {http://arxiv.org/abs/arXiv:1609.00048}
  {arXiv:1609.00048} \BibitemShut {NoStop}%
\bibitem [{\citenamefont {Eisert}\ and\ \citenamefont
  {Osborne}(2006)}]{Eisert2006QuenchEntanglement}%
  \BibitemOpen
  \bibfield  {author} {\bibinfo {author} {\bibfnamefont {J.}~\bibnamefont
  {Eisert}}\ and\ \bibinfo {author} {\bibfnamefont {T.~J.}\ \bibnamefont
  {Osborne}},\ }\href {\doibase 10.1103/PhysRevLett.97.150404} {\bibfield
  {journal} {\bibinfo  {journal} {Phys. Rev. Lett.}\ }\textbf {\bibinfo
  {volume} {97}},\ \bibinfo {pages} {150404} (\bibinfo {year}
  {2006})}\BibitemShut {NoStop}%
\bibitem [{\citenamefont {Haegeman}\ \emph {et~al.}(2011)\citenamefont
  {Haegeman}, \citenamefont {Cirac}, \citenamefont {Osborne}, \citenamefont
  {Pi{\v{z}}orn}, \citenamefont {Verschelde},\ and\ \citenamefont
  {Verstraete}}]{Haegeman2011TDVP}%
  \BibitemOpen
  \bibfield  {author} {\bibinfo {author} {\bibfnamefont {J.}~\bibnamefont
  {Haegeman}}, \bibinfo {author} {\bibfnamefont {J.~I.}\ \bibnamefont {Cirac}},
  \bibinfo {author} {\bibfnamefont {T.~J.}\ \bibnamefont {Osborne}}, \bibinfo
  {author} {\bibfnamefont {I.}~\bibnamefont {Pi{\v{z}}orn}}, \bibinfo {author}
  {\bibfnamefont {H.}~\bibnamefont {Verschelde}}, \ and\ \bibinfo {author}
  {\bibfnamefont {F.}~\bibnamefont {Verstraete}},\ }\href {\doibase
  10.1103/physrevlett.107.070601} {\bibfield  {journal} {\bibinfo  {journal}
  {Physical Review Letters}\ }\textbf {\bibinfo {volume} {107}},\ \bibinfo
  {pages} {070601} (\bibinfo {year} {2011})}\BibitemShut {NoStop}%
\bibitem [{\citenamefont {Kibble}(1976)}]{Kibble1976Cosmic}%
  \BibitemOpen
  \bibfield  {author} {\bibinfo {author} {\bibfnamefont {T.~W.~B.}\
  \bibnamefont {Kibble}},\ }\href {http://stacks.iop.org/0305-4470/9/i=8/a=029}
  {\bibfield  {journal} {\bibinfo  {journal} {Journal of Physics A:
  Mathematical and General}\ }\textbf {\bibinfo {volume} {9}},\ \bibinfo
  {pages} {1387} (\bibinfo {year} {1976})}\BibitemShut {NoStop}%
\bibitem [{\citenamefont {Zurek}(1985)}]{Zurek1985Cosmological}%
  \BibitemOpen
  \bibfield  {author} {\bibinfo {author} {\bibfnamefont {W.~H.}\ \bibnamefont
  {Zurek}},\ }\href@noop {} {\bibfield  {journal} {\bibinfo  {journal}
  {Nature}\ }\textbf {\bibinfo {volume} {317}},\ \bibinfo {pages} {505}
  (\bibinfo {year} {1985})}\BibitemShut {NoStop}%
\bibitem [{\citenamefont {White}(2005)}]{White2005SingleCenterDMRG}%
  \BibitemOpen
  \bibfield  {author} {\bibinfo {author} {\bibfnamefont {S.~R.}\ \bibnamefont
  {White}},\ }\href {\doibase 10.1103/physrevb.72.180403} {\bibfield  {journal}
  {\bibinfo  {journal} {Physical Review B}\ }\textbf {\bibinfo {volume} {72}},\
  \bibinfo {pages} {180403} (\bibinfo {year} {2005})}\BibitemShut {NoStop}%
\bibitem [{\citenamefont {Hubig}\ \emph {et~al.}(2015)\citenamefont {Hubig},
  \citenamefont {McCulloch}, \citenamefont {Schollw{\"o}ck},\ and\
  \citenamefont {Wolf}}]{Hubig2015DMRGSubspaceExpansion}%
  \BibitemOpen
  \bibfield  {author} {\bibinfo {author} {\bibfnamefont {C.}~\bibnamefont
  {Hubig}}, \bibinfo {author} {\bibfnamefont {I.~P.}\ \bibnamefont
  {McCulloch}}, \bibinfo {author} {\bibfnamefont {U.}~\bibnamefont
  {Schollw{\"o}ck}}, \ and\ \bibinfo {author} {\bibfnamefont {F.~A.}\
  \bibnamefont {Wolf}},\ }\href {\doibase 10.1103/physrevb.91.155115}
  {\bibfield  {journal} {\bibinfo  {journal} {Physical Review B}\ }\textbf
  {\bibinfo {volume} {91}},\ \bibinfo {pages} {155115} (\bibinfo {year}
  {2015})}\BibitemShut {NoStop}%
\end{thebibliography}
%

\end{document}